\documentclass[aps,prc,twocolumn,superscriptaddress,showpacs,floatfix]{revtex4-2}

\usepackage{color}
\usepackage{subfigure}
\usepackage{amsmath,epsfig}
\usepackage{graphicx}
\usepackage{epstopdf}
\usepackage{dcolumn}
\usepackage{bm}
\usepackage{multirow}
\usepackage{hhline}
\usepackage{afterpage}
\usepackage{hyperref} 
\usepackage{float}
\usepackage{booktabs}

\newcommand{\ignore}[2]{\hspace{0in}#2}

\begin{document}
\title{Azimuthal anisotropy measurements of strange and multi-strange hadrons in U+U collisions at $\sqrt{s_{NN}} = 193$ GeV at at the BNL Relativistic Heavy Ion Collider}
\affiliation{Abilene Christian University, Abilene, Texas   79699}
\affiliation{AGH University of Science and Technology, FPACS, Cracow 30-059, Poland}
\affiliation{Alikhanov Institute for Theoretical and Experimental Physics NRC "Kurchatov Institute", Moscow 117218, Russia}
\affiliation{Argonne National Laboratory, Argonne, Illinois 60439}
\affiliation{American University of Cairo, New Cairo 11835, New Cairo, Egypt}
\affiliation{Brookhaven National Laboratory, Upton, New York 11973}
\affiliation{University of California, Berkeley, California 94720}
\affiliation{University of California, Davis, California 95616}
\affiliation{University of California, Los Angeles, California 90095}
\affiliation{University of California, Riverside, California 92521}
\affiliation{Central China Normal University, Wuhan, Hubei 430079 }
\affiliation{University of Illinois at Chicago, Chicago, Illinois 60607}
\affiliation{Creighton University, Omaha, Nebraska 68178}
\affiliation{Czech Technical University in Prague, FNSPE, Prague 115 19, Czech Republic}
\affiliation{Technische Universit\"at Darmstadt, Darmstadt 64289, Germany}
\affiliation{ELTE E\"otv\"os Lor\'and University, Budapest, Hungary H-1117}
\affiliation{Frankfurt Institute for Advanced Studies FIAS, Frankfurt 60438, Germany}
\affiliation{Fudan University, Shanghai, 200433 }
\affiliation{University of Heidelberg, Heidelberg 69120, Germany }
\affiliation{University of Houston, Houston, Texas 77204}
\affiliation{Huzhou University, Huzhou, Zhejiang  313000}
\affiliation{Indian Institute of Science Education and Research (IISER), Berhampur 760010 , India}
\affiliation{Indian Institute of Science Education and Research (IISER) Tirupati, Tirupati 517507, India}
\affiliation{Indian Institute Technology, Patna, Bihar 801106, India}
\affiliation{Indiana University, Bloomington, Indiana 47408}
\affiliation{Institute of Modern Physics, Chinese Academy of Sciences, Lanzhou, Gansu 730000 }
\affiliation{University of Jammu, Jammu 180001, India}
\affiliation{Joint Institute for Nuclear Research, Dubna 141 980, Russia}
\affiliation{Kent State University, Kent, Ohio 44242}
\affiliation{University of Kentucky, Lexington, Kentucky 40506-0055}
\affiliation{Lawrence Berkeley National Laboratory, Berkeley, California 94720}
\affiliation{Lehigh University, Bethlehem, Pennsylvania 18015}
\affiliation{Max-Planck-Institut f\"ur Physik, Munich 80805, Germany}
\affiliation{Michigan State University, East Lansing, Michigan 48824}
\affiliation{National Research Nuclear University MEPhI, Moscow 115409, Russia}
\affiliation{National Institute of Science Education and Research, HBNI, Jatni 752050, India}
\affiliation{National Cheng Kung University, Tainan 70101 }
\affiliation{Nuclear Physics Institute of the CAS, Rez 250 68, Czech Republic}
\affiliation{Ohio State University, Columbus, Ohio 43210}
\affiliation{Institute of Nuclear Physics PAN, Cracow 31-342, Poland}
\affiliation{Panjab University, Chandigarh 160014, India}
\affiliation{Pennsylvania State University, University Park, Pennsylvania 16802}
\affiliation{NRC "Kurchatov Institute", Institute of High Energy Physics, Protvino 142281, Russia}
\affiliation{Purdue University, West Lafayette, Indiana 47907}
\affiliation{Rice University, Houston, Texas 77251}
\affiliation{Rutgers University, Piscataway, New Jersey 08854}
\affiliation{Universidade de S\~ao Paulo, S\~ao Paulo, Brazil 05314-970}
\affiliation{University of Science and Technology of China, Hefei, Anhui 230026}
\affiliation{Shandong University, Qingdao, Shandong 266237}
\affiliation{Shanghai Institute of Applied Physics, Chinese Academy of Sciences, Shanghai 201800}
\affiliation{Southern Connecticut State University, New Haven, Connecticut 06515}
\affiliation{State University of New York, Stony Brook, New York 11794}
\affiliation{Instituto de Alta Investigaci\'on, Universidad de Tarapac\'a, Arica 1000000, Chile}
\affiliation{Temple University, Philadelphia, Pennsylvania 19122}
\affiliation{Texas A\&M University, College Station, Texas 77843}
\affiliation{University of Texas, Austin, Texas 78712}
\affiliation{Tsinghua University, Beijing 100084}
\affiliation{University of Tsukuba, Tsukuba, Ibaraki 305-8571, Japan}
\affiliation{United States Naval Academy, Annapolis, Maryland 21402}
\affiliation{Valparaiso University, Valparaiso, Indiana 46383}
\affiliation{Variable Energy Cyclotron Centre, Kolkata 700064, India}
\affiliation{Warsaw University of Technology, Warsaw 00-661, Poland}
\affiliation{Wayne State University, Detroit, Michigan 48201}
\affiliation{Yale University, New Haven, Connecticut 06520}

\author{M.~S.~Abdallah}\affiliation{American University of Cairo, New Cairo 11835, New Cairo, Egypt}
\author{J.~Adam}\affiliation{Brookhaven National Laboratory, Upton, New York 11973}
\author{L.~Adamczyk}\affiliation{AGH University of Science and Technology, FPACS, Cracow 30-059, Poland}
\author{J.~R.~Adams}\affiliation{Ohio State University, Columbus, Ohio 43210}
\author{J.~K.~Adkins}\affiliation{University of Kentucky, Lexington, Kentucky 40506-0055}
\author{G.~Agakishiev}\affiliation{Joint Institute for Nuclear Research, Dubna 141 980, Russia}
\author{I.~Aggarwal}\affiliation{Panjab University, Chandigarh 160014, India}
\author{M.~M.~Aggarwal}\affiliation{Panjab University, Chandigarh 160014, India}
\author{Z.~Ahammed}\affiliation{Variable Energy Cyclotron Centre, Kolkata 700064, India}
\author{I.~Alekseev}\affiliation{Alikhanov Institute for Theoretical and Experimental Physics NRC "Kurchatov Institute", Moscow 117218, Russia}\affiliation{National Research Nuclear University MEPhI, Moscow 115409, Russia}
% ...

\collaboration{STAR Collaboration}\noaffiliation

\begin{abstract}
We present systematic measurements of azimuthal anisotropy for strange and multistrange hadrons ($K^{0}_{s}$, $\Lambda$, $\Xi$, and $\Omega$) and $\phi$ mesons at midrapidity ($|y| <$ 1.0) in collisions of U + U nuclei at $\sqrt{s_{NN}} = 193$~GeV, recorded by the STAR detector at the Relativistic Heavy Ion Collider. Transverse momentum ($p_{\text{T}}$) dependence of flow coefficients ($v_{2}$, $v_{3}$, and $v_{4}$) is presented for minimum bias collisions and three different centrality intervals. Number of constituent quark scaling of the measured flow coefficients in U + U collisions is discussed. We also present the ratio of $v_{n}$ scaled by the participant eccentricity ($\varepsilon_{n}\left\lbrace 2 \right\rbrace$) to explore system size dependence and collectivity in U + U collisions. The magnitude of $v_{2}/\varepsilon_{2}$ is found to be smaller in U + U collisions than that in central Au + Au collisions contradicting naive eccentricity scaling. Furthermore, the ratios between various flow harmonics ($v_{3}/v_{2}^{3/2}$, $v_{4}/v_{2}^{4/2}$) are studied and compared with hydrodynamic and transport model calculations.
\end{abstract}
	
\maketitle

\section{INTRODUCTION}
Under extreme conditions of high temperature and energy densities, quantum chromodynamics (QCD) hadronic matter undergoes a phase transition into a state of matter consisting of deconfined quarks and gluons, known as the Quark-Gluon Plasma (QGP)~\cite{qgp1,qgp2,qgp3}. Experiments at the Relativistic Heavy-Ion Collider (RHIC) and Large Hadron Collider (LHC) facilities are designed to study the properties of such a deconfined state of partonic matter created in relativistic heavy-ion collisions.

The measurements of particle production in momentum space relative to the reaction plane at RHIC~\cite{HICs1,HICs2,HICs3,HICs4} and LHC~\cite{HICl1,HICl2,HICl3} have demonstrated collective behavior of the partonic matter produced in relativistic heavy-ion collisions. Anisotropies in particle production relative to the reaction plane arise from the azimuthal asymmetry of the initial overlap region in heavy-ion collisions with non-zero impact parameter~\cite{flowt1,flowt2,flowt3} and fluctuations of nucleon positions in heavy-ion collisions~\cite{flowt4,flowt5,flowt6}. The initial spatial anisotropies are transformed into momentum space anisotropies through the interactions among quarks and gluons in the early stages prior to hadronization, and among produced hadrons at the later stages. The resulting final state momentum-space anisotropy, also referred to as anisotropic flow, can be measured using the Fourier expansion of azimuthal angle ($\varphi$) dependence of produced particles~\cite{flowm1,flowm2} according to the equation
\begin{equation}
\label{eq:invyield}
E\frac{d^{3}N}{dp^{3}} = \frac{d^{2}N}{2\pi p_{\text{T}} dp_{\text{T}} dy} \left( 1 + 2 \sum_{n=1}^{\infty} v_{n} \cos\left[n\left(\varphi - \Psi_{FP}\right)\right] \right),
\end{equation}     
where $E$, $p$, $p_{\text{T}}$, $\varphi$, and $y$ are the energy, total momentum, transverse momentum, azimuthal angle, and rapidity of the emitted particles, respectively, and $\Psi_{FP}$ is the azimuthal angle of the $n^{th}$-order flow symmetry plane (FP). For $n =$ 1, 2 the FP is related to the reaction plane (RP) through symmetry. The $\Psi_{RP}$ is defined as the plane formed by the beam axis and impact parameter vector between the centers of the two colliding nuclei. The $n^{th}$-order Fourier coefficients are defined as $v_{n} = \left\langle\cos\left[n(\varphi-\Psi_{FP})\right]\right\rangle$, where the angular brackets denote averaging over all particles in all events. 

The $2^{nd}$-order Fourier coefficient, $v_{2}$, known as elliptic flow, has played a crucial role in determining that the QGP formed at RHIC has a small shear viscosity to entropy density ratio ($\eta/s$) close to the quantum limit for a strongly-coupled quantum fluid~\cite{sQGP1}. Elliptic flow, due to its self-quenching nature, is particularly sensitive to the properties of the medium in the initial stages of heavy-ion collisions~\cite{flowe1,flowe2,flowe3}. However, hadronic re-scattering in later stages of the system evolution may reduce the sensitivity of $v_{2}$ to the early stages~\cite{med1,med2}. The $\phi$-meson and multi-strange hadrons ($\Xi$ and $\Omega$) are expected to have small hadronic interaction cross-sections compared to non-strange hadrons~\cite{phic1,phic2}. Systematic study of the transverse momentum spectra of these multi-strange hadrons also indicates that their freeze-out temperatures are close to the QGP phase transition temperature $T_{c}$ predicted by lattice QCD calculations~\cite{HICs4,mhadfo1,mhadfo2,mhadpro1}. Therefore, the observed anisotropic flow of multi-strange hadrons primarily reflects the flow from the partonic stage in heavy-ion collisions~\cite{mhadflow1,mhadflow2}.

The $3^{rd}$- and $4^{th}$-order Fourier coefficients are termed triangular flow $v_{3}$ and quadrangular flow $v_{4}$, respectively. The $3^{rd}$ harmonic ($v_{3}$) was assumed to be zero due to the symmetry in the initial overlap geometry in early days until it was realized that initial geometry fluctuations could break this symmetry and generate a finite $v_{3}$~\cite{flowm3,flowm4,flowm5}. As a consequence, there are non-zero odd harmonics present in the initial state of the colliding system~\cite{flowm6}. Because $v_{3}$ originates from these fluctuations, its direction is not correlated with the RP of the event~\cite{flowm3,v3flow1,v3flow2}.The $4^{th}$ harmonic ($v_{4}$) is originated both by these same fluctuations and by the non-linear hydrodynamic response of the medium~\cite{v4flowt1,v4flowt2}. 

It has been suggested in Refs.~\cite{flowm4,hflow2} that transverse momentum dependence of the higher-order flow coefficients ($n \geq$ 3) is a more sensitive probe for $\eta/s$, the initial state geometry, and fluctuations than the elliptic flow. Different flow harmonics depend differently on the shear viscosity of the system ($\eta/s$) and the details of the initial conditions which are determined by the dynamics and fluctuations in the colliding system. In Ref.~\cite{vnetas1} the authors used a framework of event-by-event (3+1)-dimensional viscous relativistic hydrodynamics. Their results suggest that the flow harmonics $v_{n}$ strongly depend on the value of $\eta/s$. Figure 8 of Ref.~\cite{vnetas1} shows that higher-order harmonics ($n \geq$ 3) are suppressed more than the $v_{2}$ when calculations include finite $\eta/s$ compared to ideal hydrodynamics with $\eta/s = $ 0. Higher harmonics are substantially more affected by the shear viscosity than $v_{2}$ and hence are a much more sensitive probe of $\eta/s$. This behavior is expected because diffusive processes smear out finer structures corresponding to higher n more efficiently than larger scale structures~\cite{vnetas2}. Hence, studies of both the elliptic flow and higher-order flow coefficients are important to constrain the initial conditions and for the understanding of the medium created in heavy-ion collisions.

Additionally, stronger constraints on transport and hydrodynamic model calculations can be achieved by studying the azimuthal anisotropy of identified particles as a function of transverse momentum and collision centrality. Experimental results on the $p_{\text{T}}$ dependence of $v_{2}$ of identified hadrons have provided valuable insights for the medium produced in Au+Au collisions at RHIC~\cite{flowe3,mhadflow1,mhadflow2,idflowe1}. A hadron-mass dependence of $v_{2}(p_{\text{T}})$ is observed for identified hadrons in the low-$p_{\text{T}}$ region, $p_{\text{T}} \leq$ 2 GeV/$c$, which is understood to result from hydrodynamic expansion of the medium. For the intermediate-$p_{\text{T}}$ region, $2 \leq p_{\text{T}} \leq 4$ GeV/$c$, the values of $v_{2}(p_{\text{T}})$ for identified hadrons show a baryon-meson splitting, i.e. $v_{2}$ of baryons is larger than that of mesons. This observation can be explained via quark coalescence models in which partons develop flow during the partonic evolution and the hadron flow is the sum of collective flows of constituent partons. This particle formation mechanism leads to the observed number of constituent quark (NCQ) scaling of $v_{2}(p_{\text{T}})$ at RHIC. The higher order flow coefficients also exhibit similar dependence on particle mass and particle-type up to the intermediate $p_{\text{T}}$ region~\cite{hflow3}. 

Initial conditions in heavy-ion collisions determine various measured properties of the QGP medium and so must serve as input to fluid-dynamical calculations. Observation of large elliptic flow and jet quenching (strong suppression of high-$p_{\text{T}}$ particle production) indicates that the system produced in central Au+Au collisions at top RHIC energies has a dynamical behavior of an almost ideal fluid~\cite{HICs1,HICs2,HICs3,HICs4}. However, experimental measurements are limited due to uncertainties in the initial conditions of the produced medium in heavy-ion collisions~\cite{flowt6}. One way to control or vary these initial conditions is to perform collisions of Uranium nuclei which have a deformed shape. Uranium nuclei have a prolate shape~\cite{uu1}, hence there are collision configurations (called e.g. body-body collisions) in which the initial overlap region is not spherical even in central collisions. Furthermore, depending on the angles of the two colliding Uranium nuclei relative to the reaction plane, several other collision configurations of U+U collisions are possible~\cite{uu2,uu3,uu4}. Studying these different collision configurations will provide a reference for the initial conditions in models~\cite{IS1,IS2,IS3}. In particular, it has been shown that the energy density could be increased even further in U+U collisions compared to Au+Au collisions to test ideal hydrodynamic behavior of the elliptic flow~\cite{uu2}.
 
In this paper, we present the results on flow coefficients $v_{n}$ ($n =$ 2, 3, and 4) of $K^{0}_{s}$, $\phi$, $\Lambda$, $\Xi$, and $\Omega$ at mid-rapidity ($|y| <$ 1.0) in U+U collisions at $\sqrt{s_{NN}} = 193$~GeV, as measured with the STAR detector at RHIC. The flow coefficients are studied as a function of $p_{\text{T}}$ for minimum bias (0-80\% collision centrality) and three different centrality classes (0-10\%, 10-40\%, and 40-80\%) in U+U collisions. The results are compared with published results from Au+Au collisions at $\sqrt{s_{NN}} = 200$~GeV. NCQ scaling properties of $v_{n}$ coefficients with the transverse kinetic energy are reported. We investigate system size dependence and collectivity in U+U collisions through eccentricity-scaled $v_{n}$ coefficients. The ratios between various flow harmonics ($v_{3}/v_{2}^{3/2}$, $v_{4}/v_{2}^{4/2}$) are sensitive to the properties of the medium and mechanism of hadronization. We studied these ratios as a function of $p_{\text{T}}$ for $K^{0}_{s}$, $\phi$, $\Lambda$, $\Xi$, and $\Omega$ in U+U collisions. The measured flow coefficients are compared with the hydrodynamic and transport model calculations.

This paper is organized as follows. We discuss the STAR detector system, event and centrality selection, track selection, and particle identification technique in sub-sections~\ref{star}-\ref{particle_identifcation} of section~\ref{experiment}. Methods for reconstruction of $K^{0}_{s}$, $\phi$, $\Lambda$, $\Xi$, and $\Omega$ are discussed in sub-sections~\ref{part_recons} and~\ref{comb_back}. Analysis methods for the calculation of flow coefficients are presented in sub-section~\ref{flow_method}. The systematic uncertainties associated with the $v_{n}$ measurements are discussed in section~\ref{sys_error}. The results of $v_{n}$ measurements for $K^{0}_{s}$, $\phi$, $\Lambda$, $\Xi$, and $\Omega$ are presented in section~\ref{result}. We discuss $p_{\text{T}}$ and centrality dependence of $v_{n}$ coefficients, NCQ scaling, participant eccentricity scaling, ratios of $v_{n}$ coefficients and comparison to model calculations in sub-sections~\ref{ptdep}-\ref{vnratio}. Finally, we give a summary in section~\ref{summary} of the results reported in this paper. 

\section{EXPERIMENTAL SETUP and ANALYSIS}
\label{experiment}
\subsection{STAR Detector System}
\label{star}
The Solenoidal Tracker at RHIC (STAR)~\cite{RHIC1,RHIC2} is designed to measure a large number of charged particles produced over a large solid angle in central nucleus-nucleus collisions. Its major scientific goal is to study QCD under extremely high temperature and large energy densities. Its main features include high precision tracking, momentum analysis and identification of charged particles produced at mid-rapidity. A complete overview of the STAR detector and its subsystems can be found in Ref.~\cite{STARd1}.

The data used in this analysis are based on the minimum-bias trigger using the Vertex Position Detectors (VPDs)~\cite{STARvpd1} and Zero Degree Calorimeters (ZDCs)~\cite{STARzdc1}. The two VPD detectors, located at 4.24 $< |\eta| <$ 5.1 are used to define a minimum-bias trigger, which requires a coincidence between the East and West VPD~\cite{STARtrigger1}. The VPD also provides the start time of the collision and the position of the collision vertex along the beam direction. In addition to the VPDs, a pair of ZDC detectors is used to select minimum-bias triggered data~\cite{STARtrigger1}. The ZDCs are hadronic calorimeter detectors situated on both ends of the STAR detector system at a distance of 18 m from the center. These are placed very close to the beam pipe at zero degree angles ($\theta <$ 2 mrad) to measure energy deposited by the spectator neutrons in a collision. 

The main tracking device of the STAR experiment is a Time-Projection Chamber (TPC)~\cite{STARtpc1}. It is a gas detector filled with a P10 gas (90\% Ar and 10\% CH$_{4}$). The P10 gas is regulated at a pressure of 2 mbar above atmospheric pressure. The TPC has full azimuthal (2$\pi$) coverage and a uniform pseudorapidity range of $|\eta| <$ 1 in a homogeneous magnetic field of 0.5 Tesla along the beam direction (z-axis). The TPC detector provides a momentum measurement for each charged-particle track as well as particle-identification through ionization energy loss ($\langle dE/dx \rangle$) in the gas volume. It can identify and separate pion and kaon tracks up to $p_{\text{T}} \approx$ 0.8 GeV/$c$, and protons up to $p_{\text{T}} \approx$ 1.0 GeV/$c$. In addition to the TPC, a Time Of Flight (TOF)~\cite{STARvpd1,STARtof1,STARtof2} detector is placed around the outer radius of the TPC and is used to identify particles of higher momentum. It consists of Multigap Resistive Plate Chambers (MRPC) and covers a pseudorapidity range of $|\eta| <$ 0.9 with full azimuthal acceptance. The timing resolution of the TOF system with the start time from the VPD is $\sim$80 ps. In this analysis, both the TPC and TOF detectors are used for identification of charged particles.

\subsection{Event and Centrality Selection}
The results presented are obtained from U+U collision data at a center-of-mass energy $\sqrt{s_{NN}} = 193$~GeV collected by the STAR detector at RHIC in the year 2012. The nominal collision point is the location in the lab frame where two nuclei collide. For each collision, this is determined by finding the best common point from where tracks originate. A software cut on the position of primary vertex along the beam direction ($V_{z}$) requires it to be within $\pm$ 30 cm from the center of the TPC detector to ensure uniform coverage and acceptance. An additional cut on the difference between the $V_{z}$ positions determined by the charged tracks and VPD ($|V_{z} - {V_{z}}_{vpd}| <$ 3 cm) is applied to reject pile-up events i.e., events in which extra collisions are recorded by the TPC that are not associated with the triggered event. The radial vertex position in the plane transverse to the beam direction is defined as $V_{r} = \sqrt{V_{x}^{2} + V_{y}^{2}}$. A cut of $V_{r} < $ 2 cm is used to remove background from beam and beam-pipe interactions. 

The collision centralities of events are classified according to fractions of the total inelastic cross section. The 0-10\% centrality interval corresponds to the most central collisions (i.e., events with a small impact parameter), while the 70-80\% interval represents peripheral collisions (i.e., events with a large impact parameter). The centrality definition is based on the measured charged particle multiplicity from the TPC within pseudorapidity $|\eta|<$ 0.5, uncorrected for detection efficiencies, whose distance of closest approach to the primary vertex (DCA) is $<$ 3 cm and number of fit points $>$ 15 out of a maximum of 45 pad rows for tracking in the TPC. This multiplicity is known as the reference multiplicity. The measured reference multiplicity distribution is compared with a Monte-Carlo Glauber Model~\cite{glauber} to extract the centrality of an event as in Ref.~\cite{centrality1}. We present results only with the fraction up to 80\% due to severe trigger inefficiencies beyond the 80\% cutoff.

After applying the event and centrality selection, a total of $\sim270 \times 10^{6}$ good minimum-bias events are analyzed for the results presented in this paper.

\subsection{Eccentricity from Glauber MC model}
The $n^{th}$-order participant eccentricity ($\varepsilon_{n}$) is given by~\cite{flowm4},
\begin{equation}
\varepsilon_{n} = \frac{\sqrt{\left\langle r^{n} \cos(n\phi_{part})\right\rangle^{2} + \left\langle r^{n} \sin(n\phi_{part})\right\rangle^{2}}}{\left\langle r^{n} \right\rangle}
\end{equation}
where $r$ and $\phi_{part}$ represent the positions of participating nucleons in the polar coordinate system shifted to the center of mass of the participating nucleons, and $n$ is the order of eccentricity. The angular bracket $\left\langle\right\rangle$ denotes an average over the participant nucleons in each event. The root mean square participant eccentricity is defined as $\varepsilon_{n}\left\lbrace 2\right\rbrace = \sqrt{\left\langle\left\langle \varepsilon_{n}^{2} \right\rangle\right\rangle}$. The double angular bracket $\left\langle\left\langle\right\rangle\right\rangle$ denotes an average over the event ensemble. The values of $\varepsilon_{n}\left\lbrace 2\right\rbrace$ for different centrality intervals in U+U collisions at $\sqrt{s_{NN}}$ = 193 GeV, calculated using the MC Glauber model as in Ref.~\cite{phobos_glauber1,phobos_glauber2}, are shown in Table~\ref{tab:ecc}. The centrality selection in the MC Glauber model is based on charged particle multiplicity calculated using the two-component model with the number of participants ($N_{part}$) and number of binary nucleon-nucleon collisions ($N_{coll}$). Deformation of the U nuclei has been taken into account in the MC Glauber model while calculating the participant eccentricities. We will represent $\varepsilon_{n}\left\lbrace 2\right\rbrace$ with the symbol $\varepsilon_{n}$ throughout the paper. 

\begin{table}[!htbp]
\vspace{-0.1in}
\centering
\caption{Root mean squared participant eccentricities for various centrality intervals in U+U collisions at $\sqrt{s_{NN}}$ = 193 GeV. The errors represent statistical and systematic uncertainties added in quadrature.}
\label{tab:ecc}
\vspace{0.07in}
\begin{tabular}{cccc}
\hline \rule{0pt}{14pt} \\[-2.5ex]
   harmonics & 0-10\% & 10-40\% & 40-80\% \\[0.5ex]
\hline \rule{0pt}{14pt} \\[-2.5ex]
    {$\varepsilon_{2}\left\lbrace 2\right\rbrace$}	& {0.1725$\pm$0.005}  & {0.3237$\pm$0.019} & {0.5668$\pm$0.026}	 \\[0.5ex]
    {$\varepsilon_{3}\left\lbrace 2\right\rbrace$}	& {0.1171$\pm$0.003}  & {0.2094$\pm$0.010} & {0.4002$\pm$0.019}	 \\[0.5ex] 
    {$\varepsilon_{4}\left\lbrace 2\right\rbrace$}	& {0.1432$\pm$0.004}  & {0.2644$\pm$0.015} & {0.5131$\pm$0.022}	 \\[0.5ex] 
\hline \rule{0pt}{14pt}
\end{tabular}
\end{table} 

\subsection{Track Selection}
Charged particle tracks from the TPC within $|\eta| < $ 1.0 are used to reconstruct strange and multi-strange hadrons ($K^{0}_{s}$, $\phi$, $\Lambda$, $\Xi$, and $\Omega$). Standard track selection criteria as used in the previous published STAR papers are applied to ensure good quality of the analyzed tracks~\cite{flowe1,mhadpro1,mhadflow1,mhadflow2}. Primary charged particle tracks ($\pi^{\pm}$, $K^{\pm}$, and $p(\bar{p})$) are required to have a number of TPC fit points (nHitsFit) of at least 15 (there are 45 radial pad rows in the TPC). In addition, the number of TPC fit points compared to the number of pad rows traversed by that track (nHitsPoss) should satisfy nHitsFit/nHitsPoss $>$ 0.52 to avoid over counting due to tracks that are artificially split into two by the tracking algorithm. 
Furthermore, for the $\phi$-meson analysis the distance of closest approach (DCA) of tracks from the reconstructed primary vertex is required to be less than 3 cm to reduce the contamination of secondary tracks from weak decays. The analysis of strange and multi-strange hadrons is done within mid-rapidity $|y| <$ 1. Basic track selection criteria for the tracks used in the reconstruction of $K^{0}_{s}$, $\phi$, $\Lambda$, $\Xi$, and $\Omega$ in U+U collisions are given in Table~\ref{tab:trackcuts}. Various topological selection criteria used for the reconstruction of strange and multi-strange hadrons are listed in Tables~\ref{tab:v0topcuts} and ~\ref{tab:mstopcuts}.
  
\begin{table}[!htbp]
\centering
\caption{Track selection criteria in U+U collisions at $\sqrt{s_{NN}} = 193$ GeV.}
\label{tab:trackcuts}
\begin{tabular}{l@{\hskip 0.3in}l}
\hline \rule{0pt}{12pt} \\[-2ex]
Cut 				& 	Value 		\\[1ex]
\hline \rule{0pt}{12pt} \\[-2ex]
$|\eta|$ 			& 	$<$ 1.0 	\\[1ex]
$|y|$	 			& 	$<$ 1.0 	\\[1ex]
nHitsFit 			& 	$\geq 15$	\\[1ex]
nHitsFit/nHitsPoss 	& 	$\geq 0.52$	\\[1ex]
\hline \rule{0pt}{12pt}
\end{tabular}
\end{table}

\subsection{Particle Identification}
\label{particle_identifcation}
Identification of charged particles is carried out using the STAR TPC and TOF detectors. Identification of $\pi^{\pm}$, $K^{\pm}$, and $p(\bar{p})$ are done by measuring the specific ionization energy loss ($\left\langle dE/dx \right\rangle$) in the TPC.
\begin{figure}[!htbp]
\centering
\includegraphics[width=0.4\textwidth]{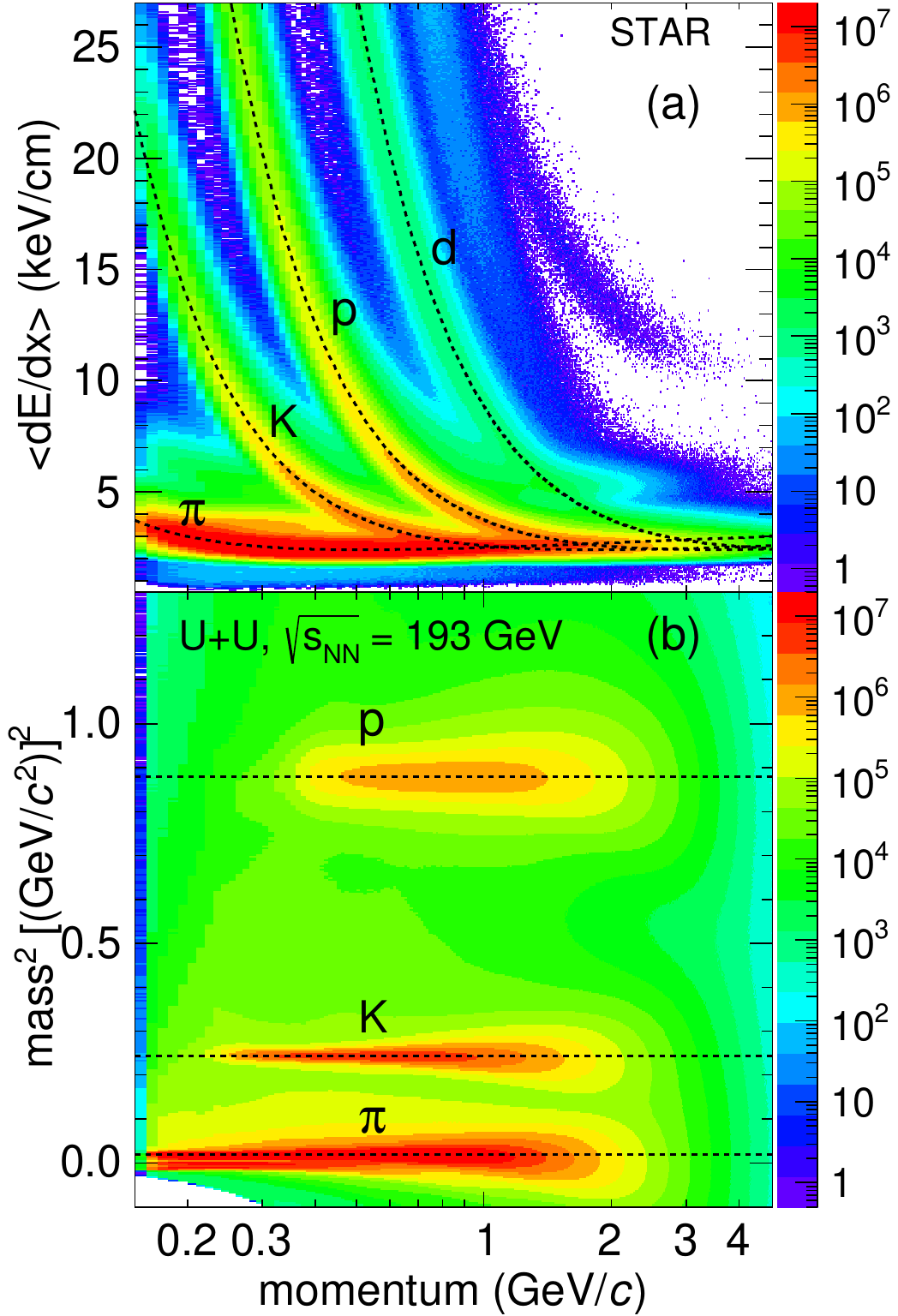}
\caption{(a) The $\left\langle dE/dx \right\rangle$ distribution of charged particles from the TPC as a function of momentum within $|\eta| <$ 1.0 for U+U collisions at $\sqrt{s_{NN}} =$ 193 GeV. The curves represent the expected mean value of $\left\langle dE/dx \right\rangle$ calculated using the Bichsel function for the corresponding particle. (b) Mass squared as a function of momentum from TOF in U+U collisions at $\sqrt{s_{NN}} =$ 193 GeV. The dashed lines represent the mass squared values from the PDG for the corresponding particle.}
\label{fig:dedxm2}
\end{figure}

Figure~\ref{fig:dedxm2} (a) shows the $\left\langle dE/dx \right\rangle$ of charged particles as a function of momentum. Different bands correspond to the measured $\left\langle dE/dx \right\rangle$ of different particle species. The dashed curves represent theoretical values predicted by the Bichsel function~\cite{bichsel}. A normalized $\left\langle dE/dx \right\rangle$, denoted as $n\sigma$, is used for the identification of $\pi^{\pm}$, $K^{\pm}$ , and $p(\bar{p})$. It is defined as,
\begin{equation}
\label{eq:nsigma}
  n\sigma_{i} = \frac{1}{R_{i}}\log\frac{\left\langle dE/dx\right\rangle_{measured}}{\left\langle dE/dx\right\rangle_{i}^{Bichsel}},
\end{equation}
where $\left\langle dE/dx \right\rangle_{i}^{Bichsel}$ is the expected $\left\langle dE/dx \right\rangle$ calculated using the Bichsel function and $R_{i}$ is the $\left\langle dE/dx \right\rangle$ resolution of the TPC for the $i^{th}$ particle species at a given momentum. As demonstrated in Fig.~\ref{fig:dedxm2} (a), the TPC detector can identify pions and kaons up to momentum of 0.8 GeV/$c$ and protons up to 1.0 GeV/$c$. At higher momentum the bands for the different particles merge together. Time of flight information of tracks from the TOF detector is therefore used to identify particles at higher momentum ranges. The time of flight ($\tau$) is the time taken by a particle to traverse the distance ($L$) from the primary vertex to the TOF detector. Velocity ($\beta = L/c\tau$) is calculated using the time of flight and track length information. The squared mass ($m^{2}$) of the particle is calculated from the velocity ($\beta$) and the corresponding momentum information from the TPC using the relation $m^{2} = p^{2}\left(1/\beta^{2} -1 \right)$. Figure~\ref{fig:dedxm2} (b) shows the mass squared ($m^{2}$) as a function of momentum. The dashed lines are the $m^{2}$ values for pions, kaons, and protons from PDG~\cite{pdg1}.

\subsection{Reconstruction of Particles}
\label{part_recons}
The particles, $K^{0}_{s}$, $\phi$, $\Lambda(\bar{\Lambda})$, $\Xi(\bar{\Xi})$, and $\Omega(\bar{\Omega})$, have short lifetimes. We reconstruct these particles through their hadronic decay channels using the invariant mass technique. Various kinematic and topological cuts are applied to reduce the combinatorial background. The decay channels used in this analysis with corresponding branching ratios are~\cite{pdg1}:
\begin{itemize}
	\item $K^{0}_{s} \rightarrow \pi^{+} + \pi^{-}$ (69.2\%)
	\item $\phi \rightarrow K^{+} + K^{-}$ (49.2\%)
	\item $\Lambda(\bar{\Lambda}) \rightarrow p + \pi^{-} (\bar{p} + \pi^{+})$ (63.9\%)
	\item $\Xi^{-}(\bar{\Xi}^{+}) \rightarrow \Lambda + \pi^{-} (\bar{\Lambda} + \pi^{+})$ (99.887\%)
	\item $\Omega^{-}(\bar{\Omega}^{+}) \rightarrow \Lambda + K^{-} (\bar{\Lambda} + K^{+})$ (67.8\%)
\end{itemize}
The charged decay daughter tracks are identified via their ionization energy loss in the STAR TPC and the time of flight obtained from the TOF detector as discussed in Sec.~\ref{particle_identifcation}.

\subsubsection{\texorpdfstring{$\phi$}{Lg}-meson reconstruction}
The $\phi$-mesons are reconstructed using the invariant mass technique through their hadronic decay channel. The $\phi$-meson decays via the strong interaction and so has a lifetime short enough that its decay position is indistinguishable from the primary vertex. Therefore, its two daughter kaons also appear to originate from the primary vertex. For this reason, primary tracks which have DCA to the primary vertex less than 3 cm, are used to reconstruct the $\phi$-mesons. Daughter kaon tracks are identified using both the TPC and TOF detectors. A criterion of $|n\sigma_{K}| <$ 2.0 is used to select kaons for $\phi$-meson reconstruction. In order to improve the particle identification at higher momentum, $m^{2}$ information from the TOF detector is used if the TOF response is available. Photon conversion electrons/positrons contaminate the $\phi$-meson candidates if they are misidentified as kaons, and contribute significantly to the residual background in the invariant mass distribution of kaon pairs ($m_{K^{+}K^{-}}$). This contribution is removed by applying a selection criteria on the dip angle $\delta$, which is defined as
\begin{equation}
\delta = \cos^{-1} \left[ \frac{p_{\text{T}1}p_{\text{T}2} + p_{z1}p_{z2}}{p_{1}p_{2}} \right],
\end{equation}

where $p_{1}$, $p_{2}$, $p_{\text{T}1}$, $p_{\text{T}2}$, $p_{z1}$, $p_{z2}$ are total, transverse, and longitudinal momenta of the two candidate tracks. The $\delta$ was required to be greater than 0.04 radians in this analysis~\cite{phimeson1,phimeson2}. The yield of $\phi$-meson candidates is obtained as a function of invariant mass $m_{K^{+}K^{-}}$ for various $p_{\text{T}}$ intervals using all possible $K^{\pm}$ pairs in an event. 
	
\subsubsection{\texorpdfstring{$K^{0}_{s}$ and $\Lambda(\bar{\Lambda}$)}{Lg} reconstruction}
$K^{0}_{s}$ and $\Lambda(\bar{\Lambda})$ are reconstructed using the neutral $V^{0}$ topological reconstruction technique. The $K^{0}_{s}$ and $\Lambda(\bar{\Lambda})$ decay, via the weak interaction, into two oppositely-charged daughter particles at a secondary vertex, a small distance away from the primary vertex (PV). The two daughter particles form a V shaped decay topology, hence $K^{0}_{s}$ and $\Lambda(\bar{\Lambda})$ are called $V^{0}$s. Reconstruction of $K^{0}_{s}$ and $\Lambda(\bar{\Lambda})$ is done by identifying the secondary vertices employing various $V^{0}$ topological selection criteria. The decay daughter tracks are identified using the TPC and TOF detectors in the same way as described in Sec.~\ref{particle_identifcation}. A criterion of $|n\sigma_{\pi,p}| <$ 3.0 is used to select daughter pions and protons. In addition, a selection criterion on mass squared ($m^{2}$) is used whenever the matched track TOF information is available to identify pions and protons. After applying basic selection criteria for daughter tracks as given in Table~\ref{tab:trackcuts}, $V^{0}$ topology cuts are used to reconstruct $K^{0}_{s}$ and $\Lambda(\bar{\Lambda})$. A list of $V^{0}$ topological selection criteria are given in Table~\ref{tab:v0topcuts}. These selection criteria are the same as used in Ref.~\cite{thesis_paul}.

\begin{table*}[!htbp]
\centering
\caption{$V^{0}$ topology selection criteria for $K^{0}_{s}$ and $\Lambda(\bar{\Lambda})$ in U+U collisions at $\sqrt{s_{NN}} = 193$ GeV.}
\label{tab:v0topcuts}
\begin{tabular}{c|ccccc|ccccc}
\hline \rule{0pt}{14pt}
       					  					&\multicolumn{4}{c}{$K_{s}^{0}$} &  &\multicolumn{4}{c}{$\Lambda(\bar{\Lambda})$} \\[1ex]
	$p_{\text{T}}$ (GeV/$c$)		  					&& {$<$ 2.0}	&& {$\geq$ 2.0}  &	&&{$<$ 2.0}    && {$\geq$ 2.0} &  	 \\[1ex]
\hline \rule{0pt}{14pt}
	DCA of $V^{0}$ to primary vertex (cm)	&& {$\leq$ 0.7} && {$\leq$ 0.8}	 & 	&&{$\leq$ 0.7} && {$\leq$ 0.7} &	 \\[1ex] 	
	DCA between $V^{0}$ daughters (cm)		&& {$\leq$ 0.7} && {$\leq$ 0.8}	 &  &&{$\leq$ 0.7} && {$\leq$ 0.7} &	 \\[1ex] 
	DCA of $\pi$ to primary vertex (cm)		&& {$\geq$ 1.5} && {$\geq$ 0.35} &  &&{$\geq$ 1.0} && {$\geq$ 1.0} &	 \\[1ex] 
	DCA of $p$ to primary vertex (cm)		&& {--}		    && {--}			 &  &&{$\geq$ 0.5} && {$\geq$ 0.25}&     \\[1ex]	
    $V^{0}$ Decay Length (cm)				&& {$\geq$ 4.5} && {$\geq$ 7.0}	 &  &&{$\geq$ 4.5} && {$\geq$ 4.5} &	 \\[1ex]	
\hline
\end{tabular}
\end{table*}

\subsubsection{\texorpdfstring{$\Xi(\bar{\Xi})$ and $\Omega(\bar{\Omega})$}{Lg} reconstruction}
The multi-strange hadrons $\Xi(\bar{\Xi})$ and $\Omega(\bar{\Omega})$ decay into a charged particle ($\pi$ or $K$) and a neutral $V^{0}$ particle ($\Lambda(\bar{\Lambda})$). These multi-strange hadrons are reconstructed via decay topology as described in the previous sub-section. The decay daughter tracks are identified in the TPC and TOF detectors. The process of reconstruction of multi-strange hadrons involves finding of two secondary decay \ignore{vertex}vertices. The reconstruction is done in two steps. First, a decay vertex of a neutral $V^{0}$ candidate is found using decay kinematics. The next step is to find a matching charged pion or kaon for this candidate. Various geometric, kinematic, and topological cuts are applied to reduce the combinatorial background. The topological selection criteria for $\Xi$ and $\Omega$ reconstruction are listed in Table~\ref{tab:mstopcuts}. These selection criteria are optimized for $\Xi$ and $\Omega$ reconstruction and are taken from the published STAR paper~\cite{idflowe1}.

\begin{table*}[!htbp]
\centering
\caption{Topological selection criteria for $\Xi(\bar{\Xi})$ and $\Omega(\bar{\Omega})$ in U+U collisions at $\sqrt{s_{NN}} = 193$ GeV.}
\label{tab:mstopcuts}
\begin{tabular}{c|ccc|ccc}
\hline \rule{0pt}{14pt}
   Cut  											 && $\Xi$ 			 &  && $\Omega$ 		&	\\[1ex]
\hline \rule{0pt}{14pt}
   DCA of $\Xi$/$\Omega$ to primary vertex			 && {$\leq$ 0.5 cm}  &  && {$\leq$ 0.4 cm}	&	\\[1ex] 	
   DCA between $\Lambda$ and bachelor $\pi/K$		 && {$\leq$ 0.8 cm}  &  && {$\leq$ 0.7 cm}	&	\\[1ex]
   DCA of bachelor $\pi/K$ to primary vertex		 && {$\geq$ 2.0 cm}  &  && {$\geq$ 1.0 cm}	&	\\[1ex] 
   DCA of $\Lambda$ to primary vertex				 && {$\geq$ 0.7 cm}  &  && {$\geq$ 0.4 cm}	& 	\\[1ex] 
   DCA between $\Lambda$-daughters					 && {$\leq$ 0.8 cm}  &  && {$\leq$ 0.7 cm}	&	\\[1ex] 
   DCA of $\Lambda$-daughter $\pi$ to primary vertex && {$\geq$ 2.0 cm}  &  && {$\geq$ 2.0 cm}	&   \\[1ex] 
   DCA of $\Lambda$-daughter $p$ to primary vertex 	 && {$\geq$ 0.6 cm}  &  && {$\geq$ 0.6 cm}	&	\\[1ex] 
   Decay length of $\Xi$/$\Omega$					 && {$\geq$ 4.0 cm}  &  && {$\geq$ 3.0 cm}	&	\\[1ex] 
   Decay length of $\Lambda$ 						 && {$\geq$ 5.0 cm}  &  && {$\geq$ 5.0 cm}	&	\\[1ex] 
   Mass width of $\Lambda$ 							 && {$\leq$ 6 MeV}   &  && {$\leq$ 6 MeV}   & 	\\[1ex]
\hline
\end{tabular}
\end{table*}

\subsection{Combinatorial Background Estimation}
\label{comb_back}
\begin{figure*}[!htbp]
\centering
\includegraphics[width=0.32\textwidth]{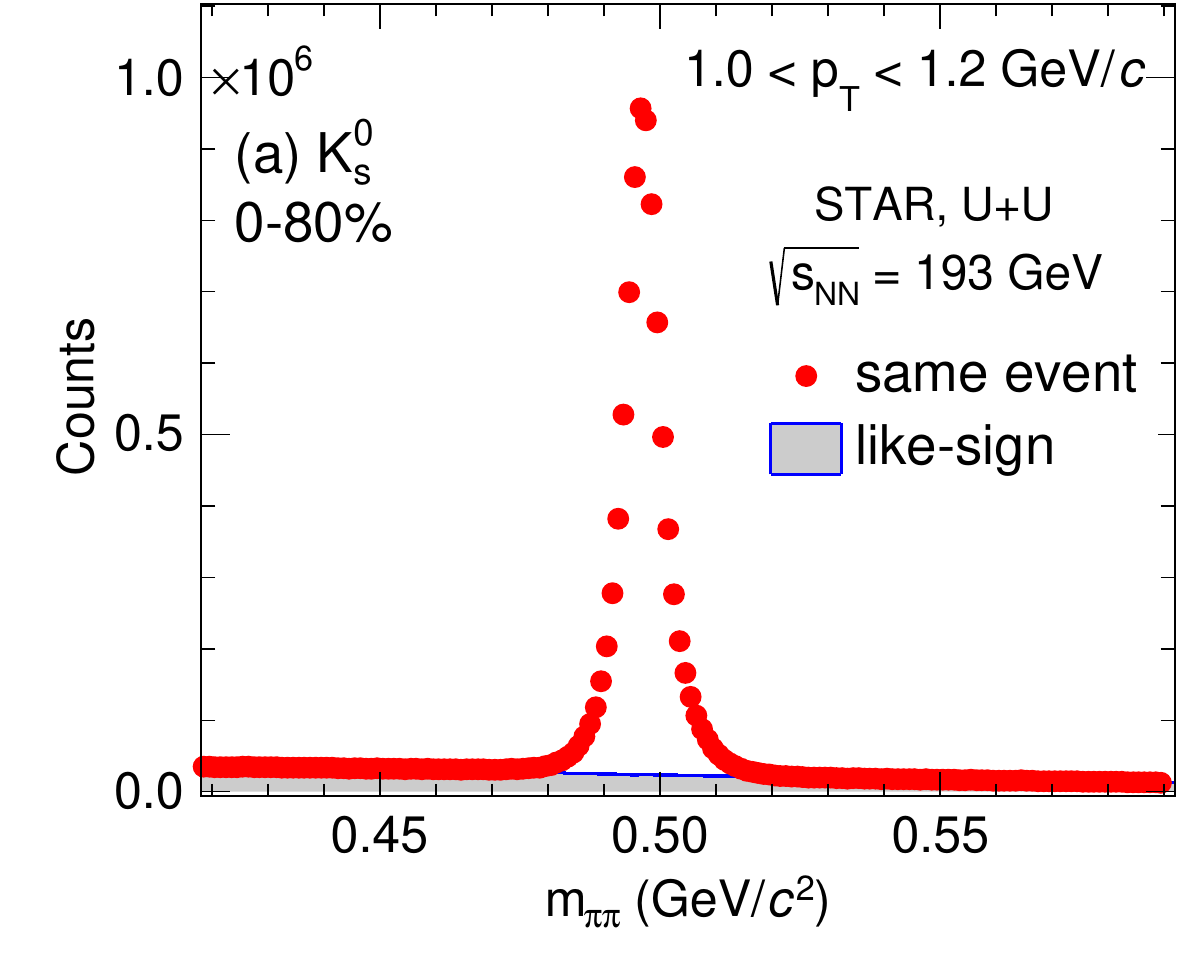} 
\includegraphics[width=0.32\textwidth]{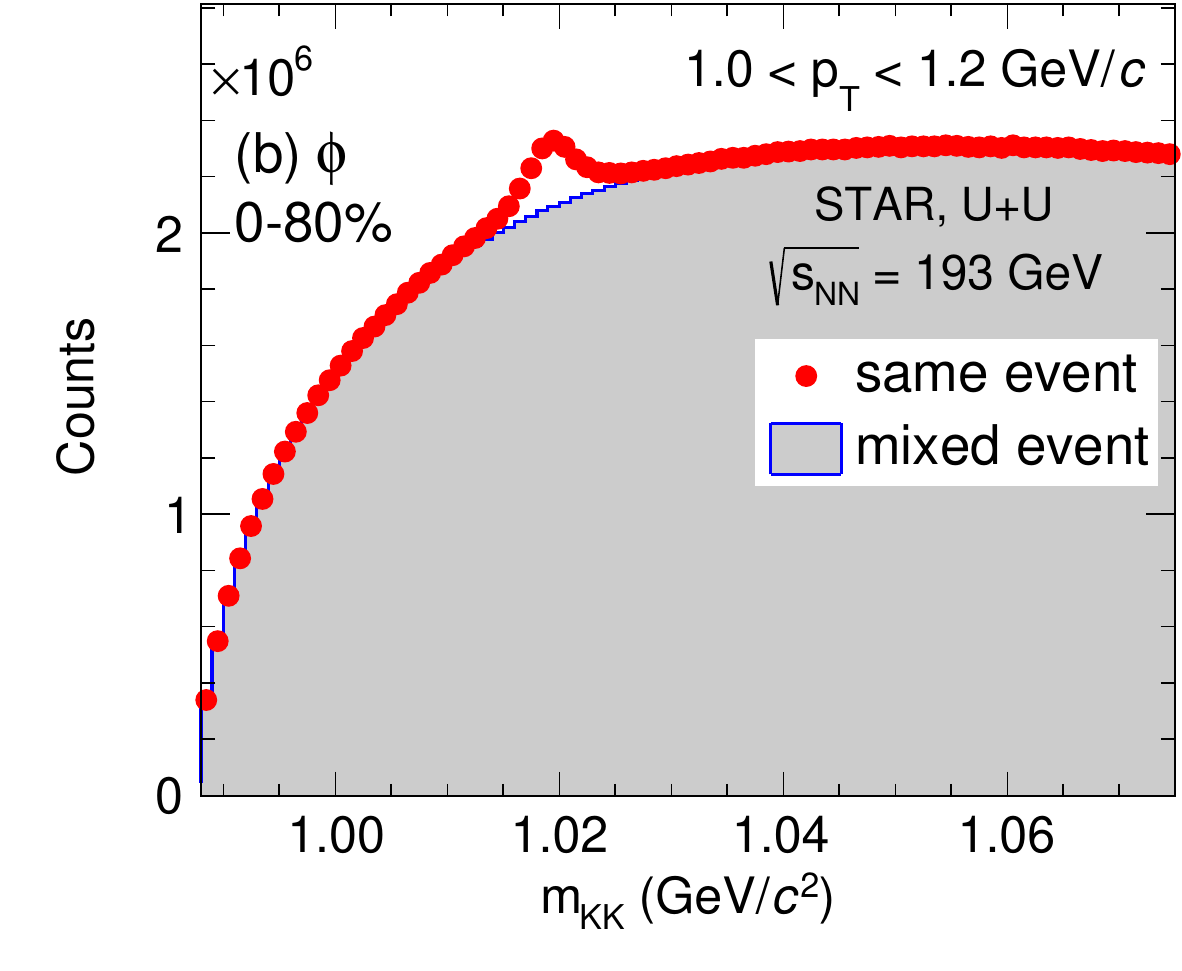}
\includegraphics[width=0.32\textwidth]{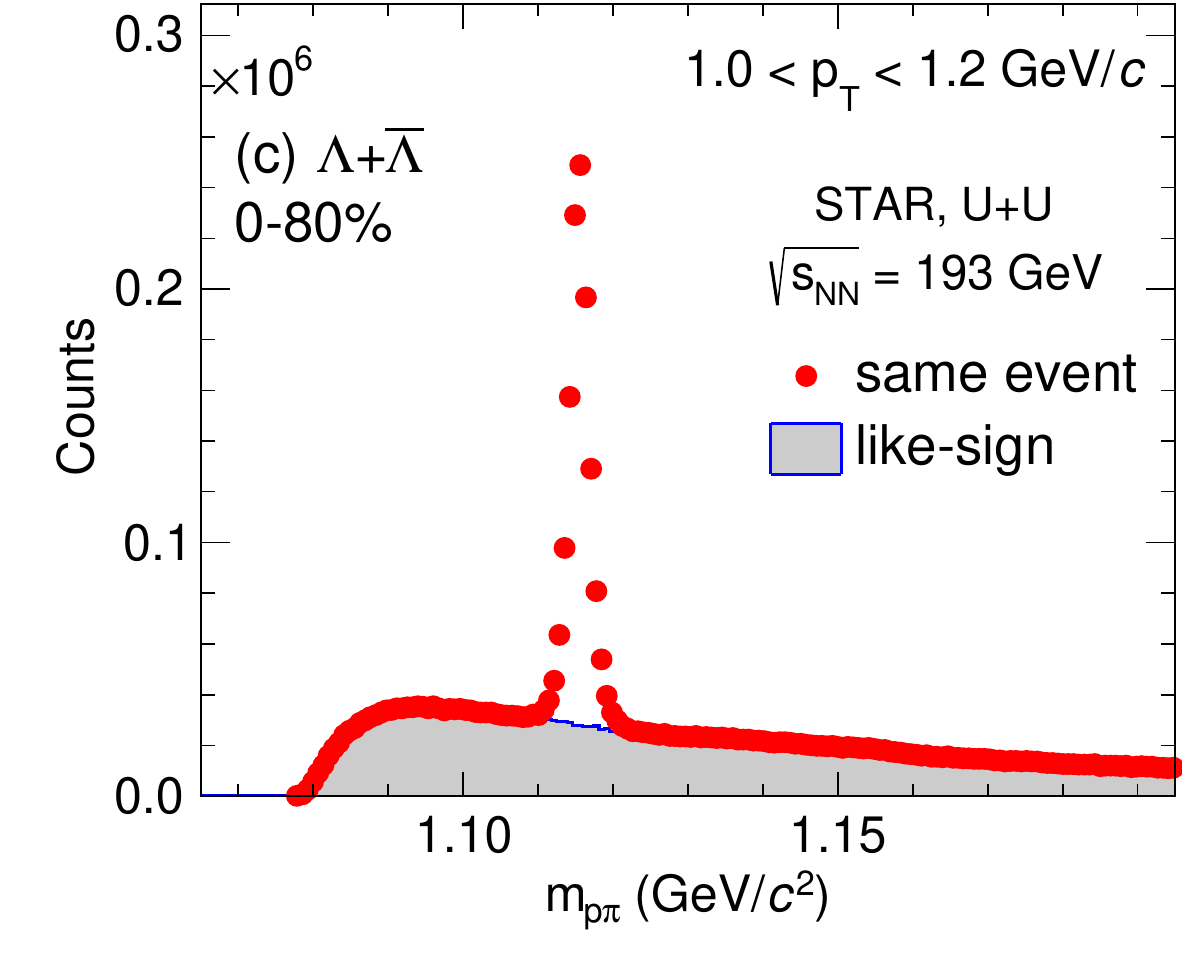} 
\includegraphics[width=0.32\textwidth]{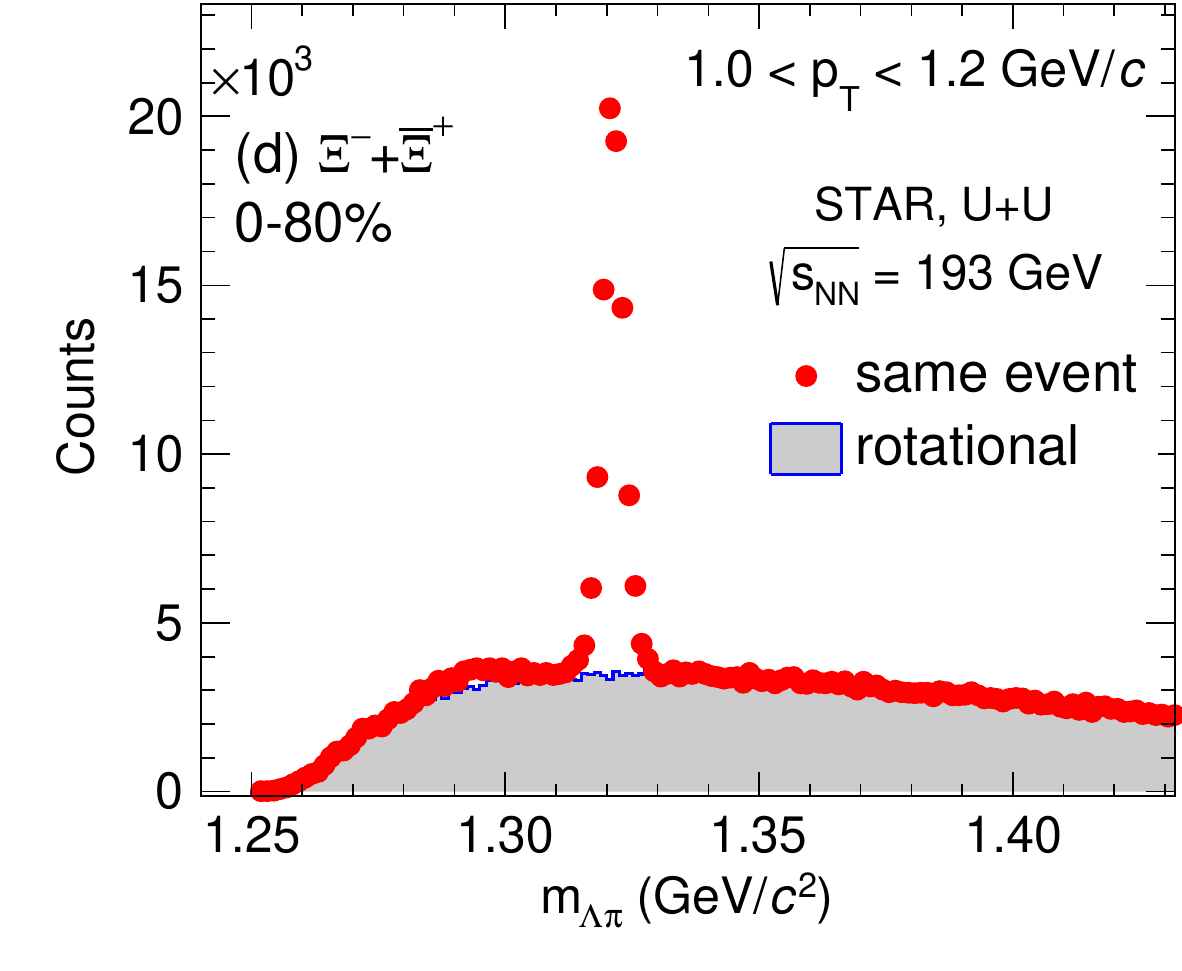}
\includegraphics[width=0.32\textwidth]{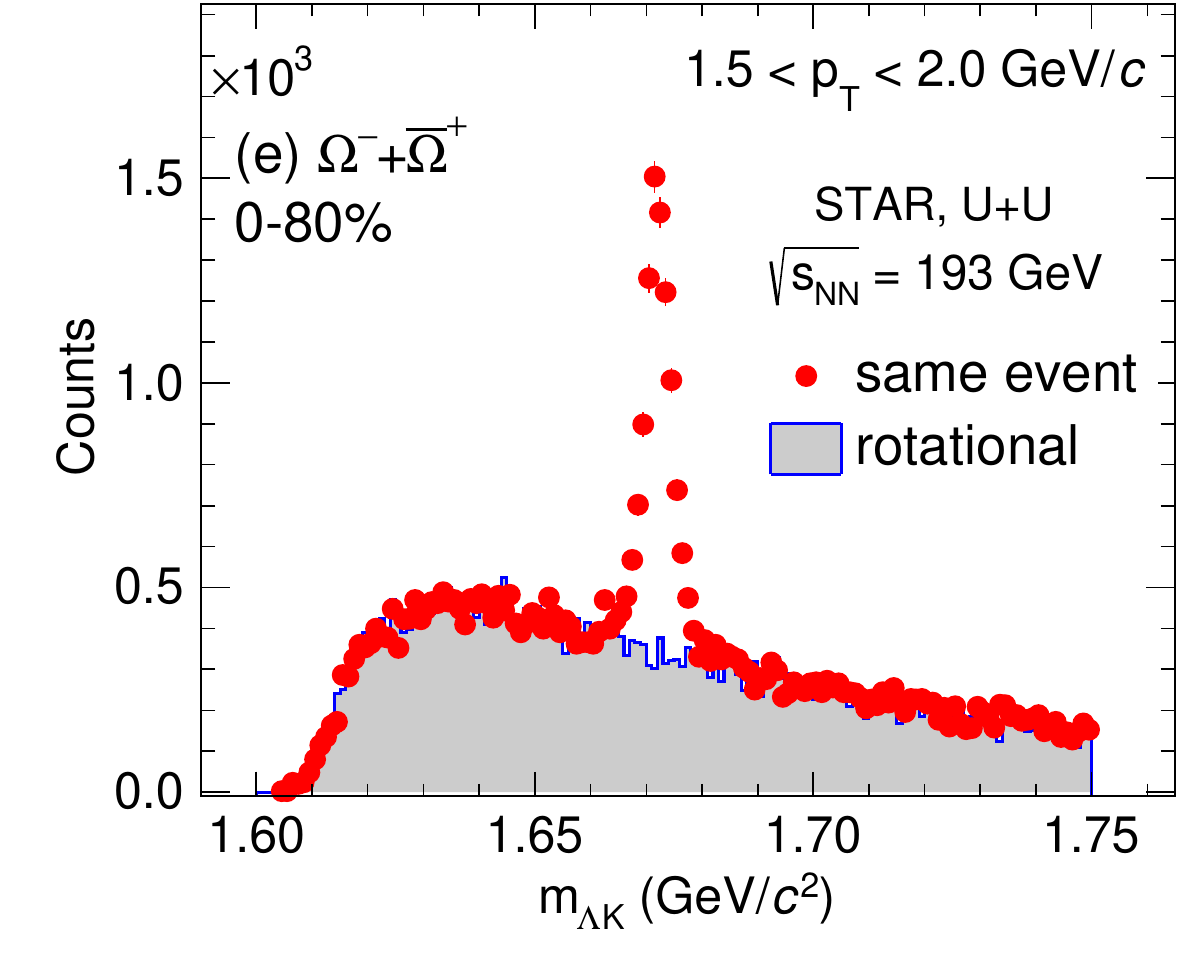}
\caption{Invariant mass distributions for (a) $K^{0}_{s}$, (b) $\phi$, (c) $\Lambda + \bar{\Lambda}$, and (d) $\Xi + \bar{\Xi}$ for 1.0 $ <p_{\text{T}} <$ 1.2 GeV/$c$, and (e) $\Omega + \bar{\Omega}$ for 1.5 $< p_{\text{T}} <$ 2.0 GeV/$c$ in minimum bias U+U collisions at $\sqrt{s_{NN}} = 193$ GeV. The grey bands are the estimated combinatorial backgrounds from mixed event technique for $\phi$, like sign technique for $K^{0}_{s}$ and $\Lambda(\bar{\Lambda})$, and rotational technique for $\Xi(\bar{\Xi})$ and $\Omega(\bar{\Omega})$. Error bars represent the statistical uncertainties.}
\label{fig:invmass}
\end{figure*}

Figure~\ref{fig:invmass} shows the invariant mass distributions after various selection cuts for (a) $K^{0}_{s}$, (b) $\phi$, (c) $\Lambda + \bar{\Lambda}$, (d) $\Xi + \bar{\Xi}$, and (e) $\Omega + \bar{\Omega}$ for a given $p_{\text{T}}$ range in minimum bias U+U collisions at $\sqrt{s_{NN}} = 193$ GeV. The measured invariant mass distributions contain both signal (S) and background (B). A clear signal peak above the combinatorial background is seen around the rest mass of the particle. The random combinatorial background is best estimated using the mixed event technique for the $\phi$-meson, the like-sign technique for $K^{0}_{s}$ and $\Lambda(\bar{\Lambda})$, and a rotational technique for $\Xi(\bar{\Xi})$ and $\Omega(\bar{\Omega})$ as described in Refs.~\cite{mhadflow1,phimeson1,ksl1}.

\subsubsection{Mixed event background}
The combinatorial background for the $\phi$-meson from uncorrelated particles is estimated using the mixed-event technique~\cite{phimeson1,phimeson2}. There are no correlations between the charged kaon tracks from different events. By mixing kaons from similar kinds of events from the same centrality class, the mixed event technique reproduces the shape of the background distribution well. Events are divided into 9 bins of centrality (from 0-5\%, 5-10\%, 10-20\% up to 70-80\%), 10 bins of 6 cm in z-vertex ($V_{z}$) between $\pm$30 cm and 5 bins of $\pi$/5 each in event-plane angle ($\psi_{n}$) between 0 to $\pi$, which makes a total of 450 event classes. For each event class, kaons from 5 different events are mixed to obtain the combinatorial background. Event mixing results in a larger number of reconstructed background candidates than the signal candidates~\cite{phimeson1,phimeson2}. Therefore, the combinatorial background is normalized to the candidate yields using an iterative method. At least four iterations are performed to scale the background distribution. The normalized background is then subtracted from the distribution of signal candidates and the resulting distribution is fitted with a Breit-Wigner function plus a $2^{nd}$-order polynomial to obtain the yield of $\phi$-mesons.

\subsubsection{Like-sign background}
The combinatorial background for $K_{s}^{0}$ and $\Lambda(\bar{\Lambda})$ is constructed using the like-sign technique~\cite{phimeson2}. In this technique, the invariant mass distribution of same-sign particle pairs from an event is obtained to reproduce the background shape. The same-sign pairs are not correlated with $K_{s}^{0}$ or $\Lambda(\bar{\Lambda})$ decays. For $K_{S}^{0}$, the like-sign background is constructed using ($\pi^{+}$+$\pi^{+}$) and ($\pi^{-}$+$\pi^{-}$) pairs. For $\Lambda(\bar{\Lambda})$, the like-sign background is constructed using ($p$+$\pi^{+}$) and ($\bar{p}$+$\pi^{-}$) pairs. The like-sign invariant mass distributions obtained for the $K_{s}^{0}$ are normalized as follows:
\begin{equation}
	N_{\pi\pi}(m) = \sqrt{N_{\pi^{+}\pi^{+}}(m) \times N_{\pi^{-}\pi^{-}}(m)},
\end{equation}
where $N$ is the number of like-sign pairs at the center of invariant mass bin $m$. The unlike-sign and the normalized like-sign invariant mass distributions are shown in Fig.~\ref{fig:invmass}(a). The normalized like-sign background is then subtracted from the unlike-sign invariant mass distribution to get the $K_{s}^{0}$ signal distribution. The resulting distribution is then used to obtain the yield of $K_{s}^{0}$ using the bin counting method as described in Ref.~\cite{ksl1,ksl2}. A similar like-sign technique is used to obtain the yield of $\Lambda(\bar{\Lambda})$.

\subsubsection{Rotational background}
For the $\Xi$ and $\Omega$, combinatorial background is constructed using the rotational method. In this method, tracks from one of the daughter type particles are rotated by $180^{\circ}$ in the transverse plane, and the resulting invariant mass distribution is used to estimate the background. Therefore, this rotated invariant mass distribution does not contain signal but reproduces the shape of the combinatorial background. In this analysis, the momentum vector of the decay daughter $\Lambda$ is rotated by $180^{\circ}$. The little residual bump at lower invariant mass in Fig.~\ref{fig:invmass}(d) is due to a $\Lambda$ decay mistakenly reconstructed as $\Xi$ topology, where the $\Lambda$ daughter proton is combined with a random pion to form a fake $\Lambda$ and the fake $\Lambda$ forms a $\Xi$ decay topology with the $\Lambda$ daughter pion. This fake $\Xi$ peak is significantly below the true $\Xi$ mass and does not affect the $\Xi$ signal extraction~\cite{mhadflow1}.

\subsection{Flow Analysis Method}
\label{flow_method}
Flow coefficients $v_{n}$ are measured using the $\eta$ sub-event plane method~\cite{flowm1,flowm2}. In this method, the event-plane angle (estimation of the reaction plane angle) for each harmonic is determined using the anisotropic flow of particles. The $n^{th}$-order event-plane angle ($\psi_{n}$) for each event is constructed using charged particle tracks from the TPC within $|\eta| <$ 1 as
\begin{equation}
\psi_{n} =\frac{1}{n}\tan^{-1} \left(\frac{Q_{ny}}{Q_{nx}}\right),
\label{eq:psin}
\end{equation}
\begin{equation}
Q_{n}\cos(n\psi_{n})\ =\ Q_{nx} = \sum\limits_{i=1}^{M} w_{i}\cos(n\phi_{i}),
\label{eq:qnx}
\end{equation}
\begin{equation}
Q_{n}\sin(n\psi_{n})\ =\ Q_{ny} = \sum\limits_{i=1}^{M} w_{i}\sin(n\phi_{i}),
\label{eq:qny}
\end{equation}
where $Q_{n}$ are the event flow vectors, $\phi_{i}$ is the azimuthal angle of the produced particle, $w_{i}$ is its weight and $M$ is the total number of particles in an event used for the flow vector calculation. In order to minimize the effects of phenomena not necessarily correlated with the event plane, called ``non-flow effects'' (jets, for example), only particles with $p_{\text{T}} <$ 2 GeV/$c$ are used in the event-plane angle calculation. The weights $w_{i}$ are set equal to $p_{\text{T}}$ up to 2 GeV/$c$ to optimize the event plane resolution. The event-plane angle distribution for an ideal detector acceptance should be isotropic in the laboratory frame. The non-uniform azimuthal detection efficiency of the TPC detector makes the reconstructed event-plane angle distribution anisotropic. This detector acceptance bias is removed by applying three methods: $\phi$-weighting, re-centering and shifting. The details of these procedures can be found in Ref.~\cite{flowm2}.  

The resolution of the event-plane angle with respect to the reaction plane is defined as~\cite{flowm2}
\begin{equation}
R = \left\langle \cos \left[n\left(\psi_{n} - \Psi_{RP}\right) \right] \right\rangle.  
\label{eq:epres}
\end{equation}
The resolution cannot be directly calculated from this equation because $\Psi_{RP}$ is unknown. Thus, the event plane resolution is estimated using the correlations between event planes calculated from two subsets of tracks, called sub-events A and B. In this analysis, we use two independent sub-events based on the pseudorapidity regions -1.0 $< \eta <$ -0.05 and 0.05 $< \eta <$ 1.0, with a gap of $\Delta\eta$ = 0.1 between the two sub-events to suppress non-flow effects. The event plane resolution for the sub-events with the assumption of only flow correlations between them is calculated by the equation~\cite{flowm2}
\begin{equation}
\left\langle \cos \left[n\left(\psi_{n} - \Psi_{RP}\right) \right] \right\rangle = \sqrt{\left\langle\cos\left[ n\left(\psi_{n}^{A} - \psi_{n}^{B}\right)\right]\right\rangle}.  
\label{eq:etasubres}
\end{equation}

\begin{figure}[!htbp]
\centering
\includegraphics[width=0.45\textwidth]{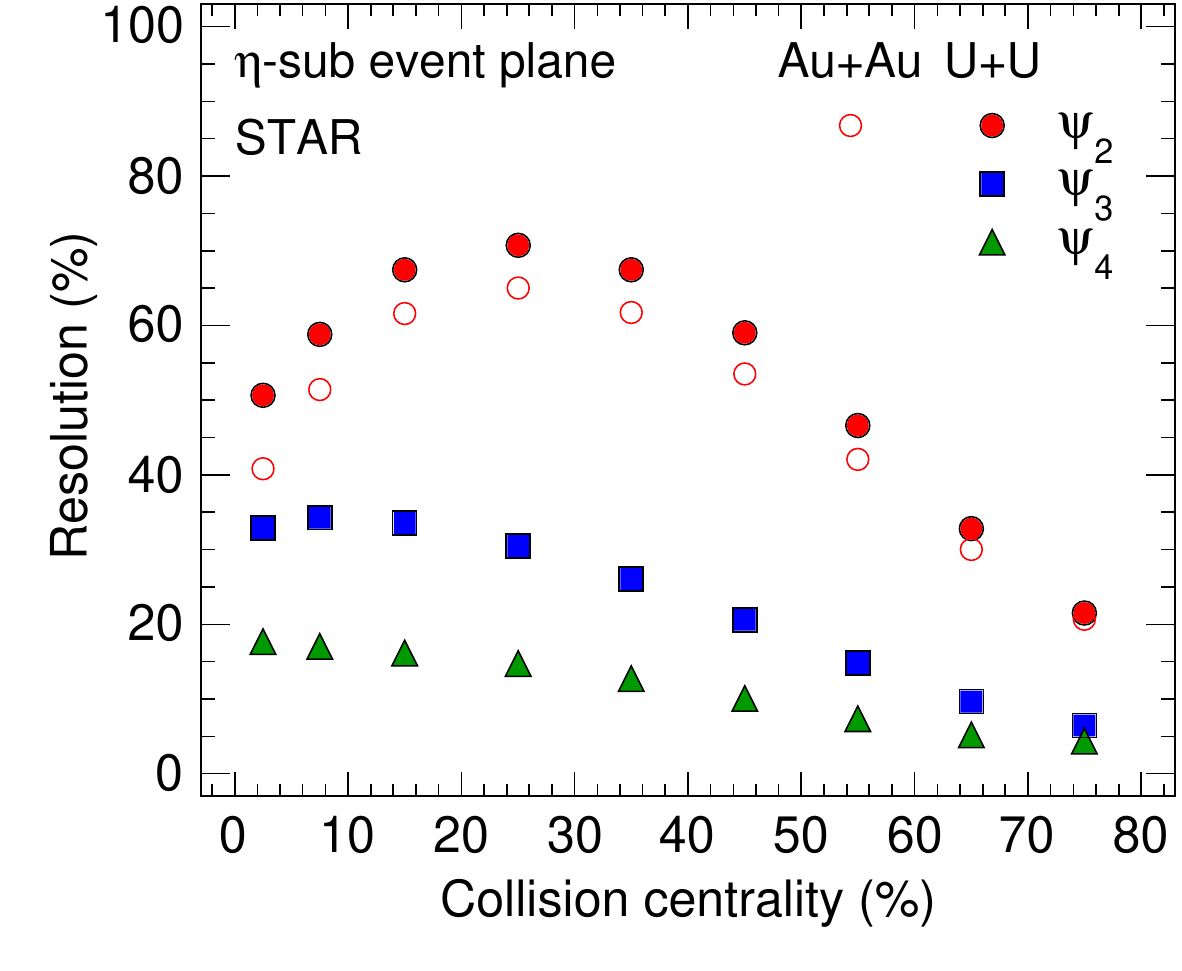}
\caption{\small{Event plane resolution as a function of centrality for $\psi_{2}$, $\psi_{3}$, and $\psi_{4}$ in U+U collisions at $\sqrt{s_{NN}}$ = 193 GeV, compared with Au+Au collisions at $\sqrt{s_{NN}}$ = 200 GeV. The statistical uncertainties are smaller than the markers.}}
\label{fig:epres}
\end{figure}

The event plane resolution depends strongly on the centrality. In this analysis, event plane resolutions are calculated for nine different centrality classes. Figure~\ref{fig:epres} shows the $\eta$ sub-event plane resolution as a function of centrality for $\psi_{2}$, $\psi_{3}$, and $\psi_{4}$ in U+U collisions at $\sqrt{s_{NN}}$ = 193 GeV. The resolutions for $\psi_{2}$ are compared with those from Au+Au collisions at $\sqrt{s_{NN}}$ = 200 GeV. The shape of event plane resolution as a function of centrality in U+U collisions is similar to that of Au+Au collisions. Resolution is higher in U+U collisions compared to Au+Au collisions likely due to higher particle multiplicity and/or higher eccentricities in U+U collisions. For combined centrality classes, i.e. 0-10\%, 10-40\%, and 40-80\%, an average resolution weighted by the raw-yield of particles is calculated. A summary of the raw-yield weighted average resolution correction factors for the combined centrality classes is shown in Table~\ref{tab:cmres}.

\begin{table}[!htbp]
\centering
\caption{Average event plane resolution for combined centrality classes in U+U collisions at $\sqrt{s_{NN}}$ = 193 GeV.}
\label{tab:cmres}
\begin{tabular}{cccccc}
\hline \rule{0pt}{14pt} \\[-2.5ex]
    \textbf{particle} & \textbf{harmonic($\psi_{n}$)} & \textbf{0-10\%} & \textbf{10-40\%} & \textbf{40-80\%} & \\[0.5ex]
\hline \rule{0pt}{14pt} \\[-2.5ex]
    {$\phi$} & 2		& {0.544}  & {0.685} & {0.468}	 \\[0.5ex] 
    {$\phi$} & 3		& {0.336}  & {0.304} & {0.151}	 \\[0.5ex] 
    {$\phi$} & 4		& {0.174}  & {0.147} & {0.078}	 \\[0.5ex] 
\hline \rule{0pt}{14pt} \\[-2.5ex]
	{$K_{s}^{0}$} & 2	& {0.548}  & {0.686} & {0.499}	 \\[0.5ex] 
	{$K_{s}^{0}$} & 3	& {0.336}  & {0.307} & {0.167}	 \\[0.5ex] 
	{$K_{s}^{0}$} & 4	& {0.174}  & {0.148} & {0.083}	 \\[0.5ex] 
\hline \rule{0pt}{14pt} \\[-2.5ex]
    {$\Lambda$} & 2		& {0.548}  & {0.686} & {0.504}	 \\[0.5ex]
	{$\Lambda$} & 3		& {0.336}  & {0.307} & {0.169}	 \\[0.5ex] 
    {$\Lambda$} & 4		& {0.174}  & {0.148} & {0.084}	 \\[0.5ex]
\hline \rule{0pt}{14pt} \\[-2.5ex]
	{$\Xi$}  & 2		& {0.544}  	& {0.685} & {0.511}	 \\[0.5ex]
    {$\Xi$}  & 3		& {0.335}  	& {0.310} & {0.171}	 \\[0.5ex]
    {$\Xi$}  & 4		& {0.174}  	& {0.150} & {0.085}	 \\[0.5ex] 
\hline \rule{0pt}{14pt} \\[-2.5ex]
	{$\Omega$} & 2		& {0.541}  	& {0.684} & {0.520}	 \\[0.5ex] 
	{$\Omega$} & 3		& {0.335}  	& {0.313} & {0.176}	 \\[0.5ex]
    {$\Omega$} & 4		& {0.174}  	& {0.151} & {0.087}	 \\[0.5ex] 
\hline \rule{0pt}{14pt}
\end{tabular}
\end{table} 

The flow coefficients $v_{n}$ are measured with respect to the estimated event-plane angle $\psi_{n}$, denoted by $v_{n}^{obs}$, as
\begin{equation}
 v_{n}^{obs} = \left\langle\cos\left[n(\varphi-\psi_{n})\right]\right\rangle.
 \label{eq:vn}
\end{equation}
The observed $v_{n}^{obs}$ coefficients are corrected by dividing the corresponding event plane resolution. Then the final $v_{n}$ coefficients are obtained as,
\begin{equation}
 v_{n} = \frac{v_{n}^{obs}}{\sqrt{\left\langle\cos\left[ n\left(\psi_{n}^{A} - \psi_{n}^{B}\right)\right]\right\rangle}}.
\label{eq:vncorr}
\end{equation}

\begin{figure*}[!htbp]
\centering
\includegraphics[width=0.3\textwidth]{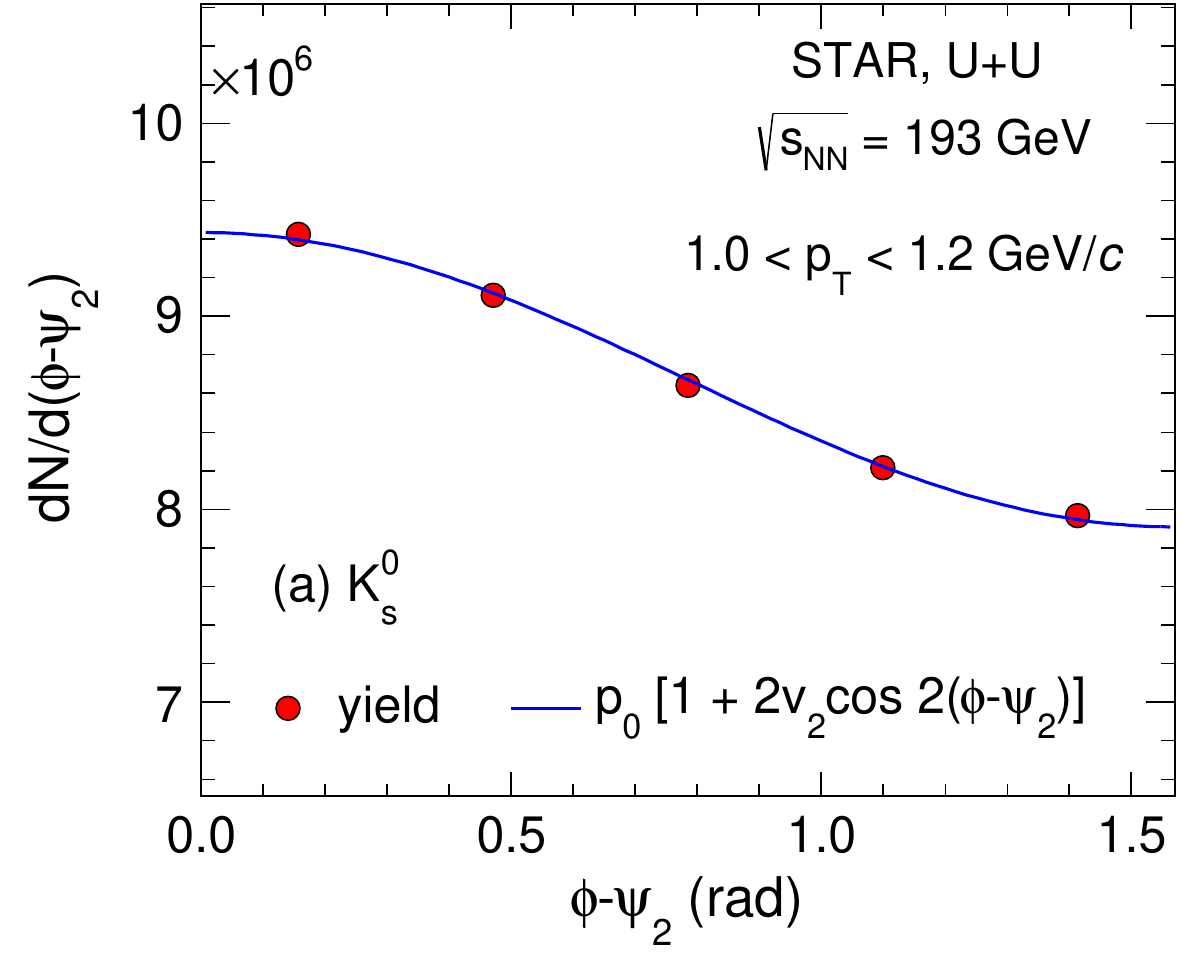} 
\includegraphics[width=0.3\textwidth]{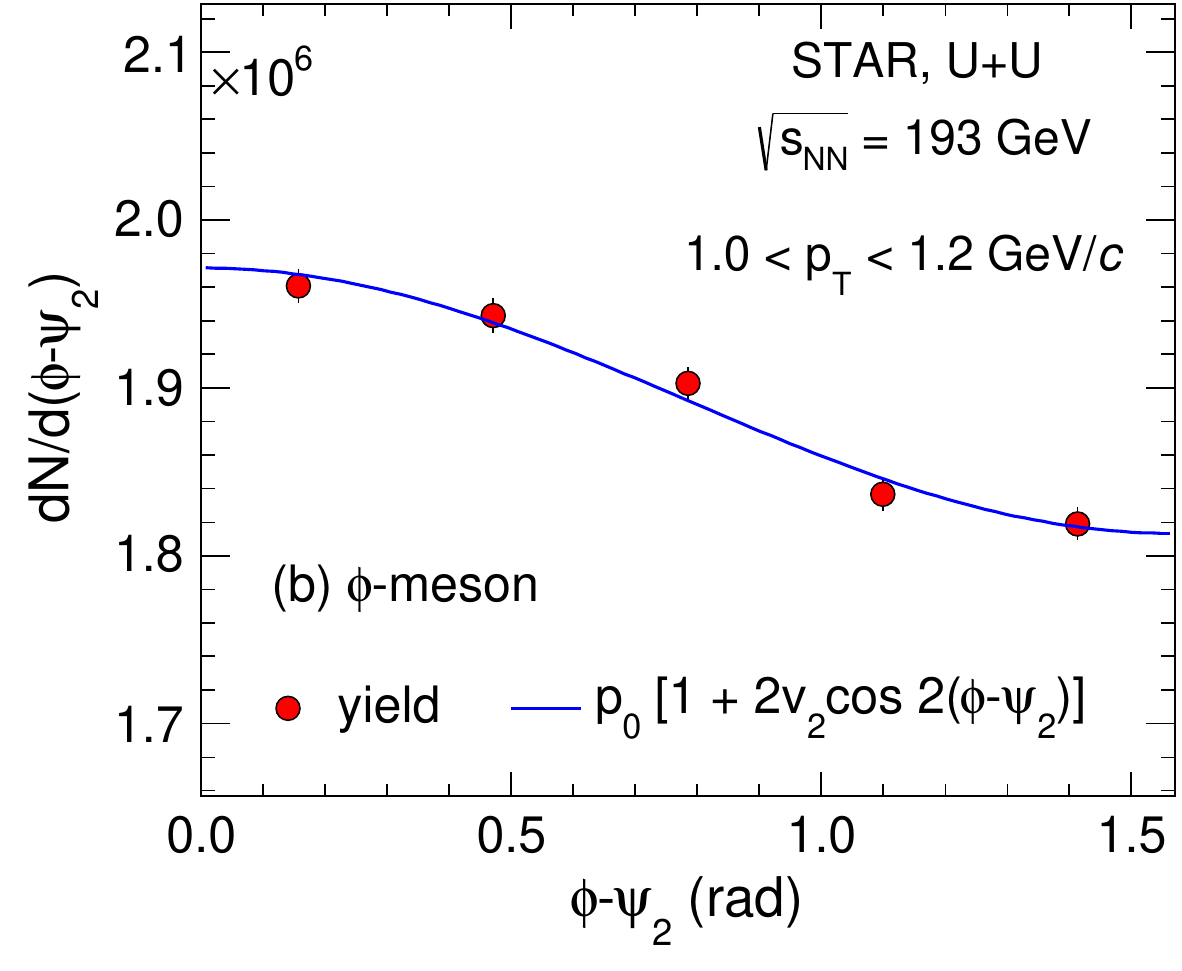}
\includegraphics[width=0.3\textwidth]{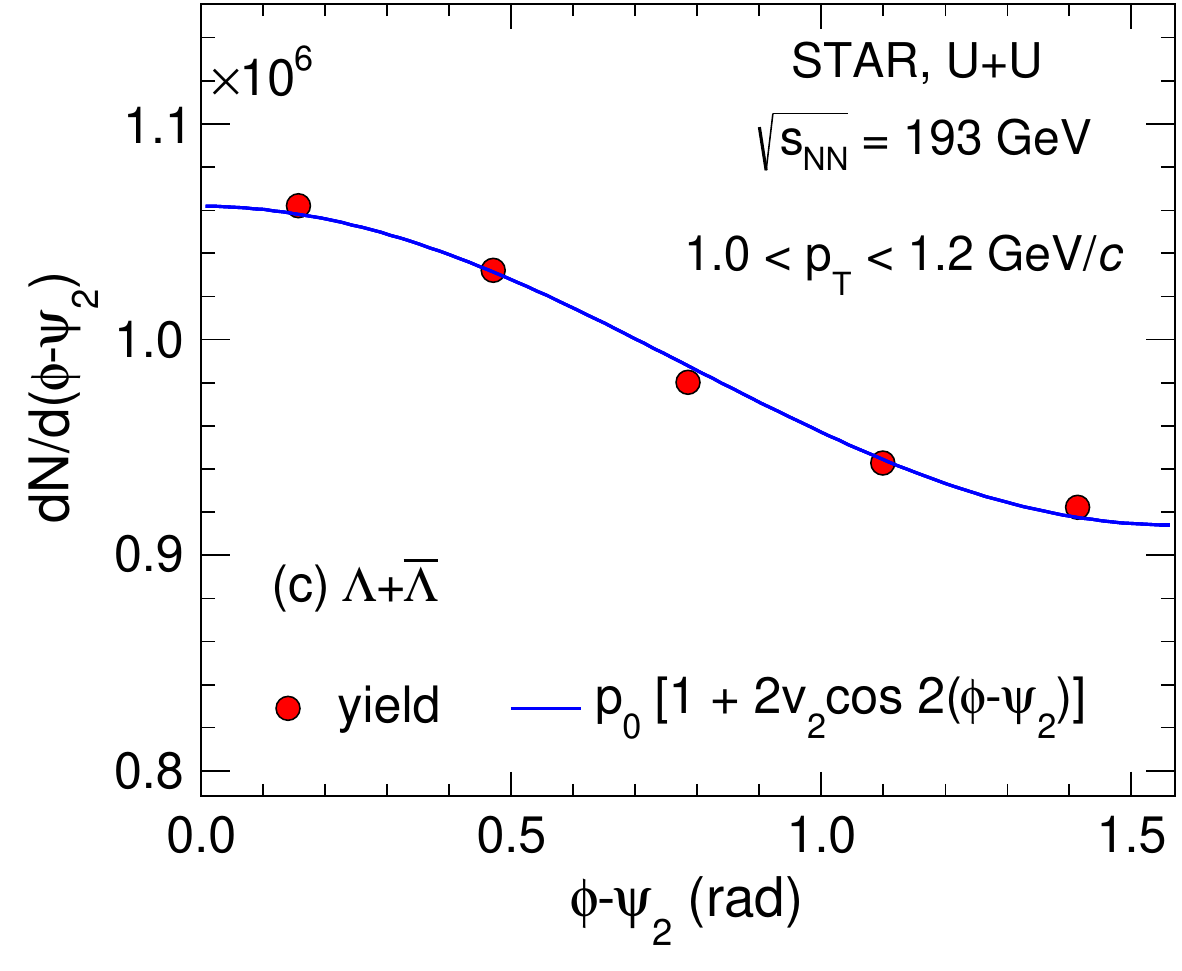} 
\includegraphics[width=0.3\textwidth]{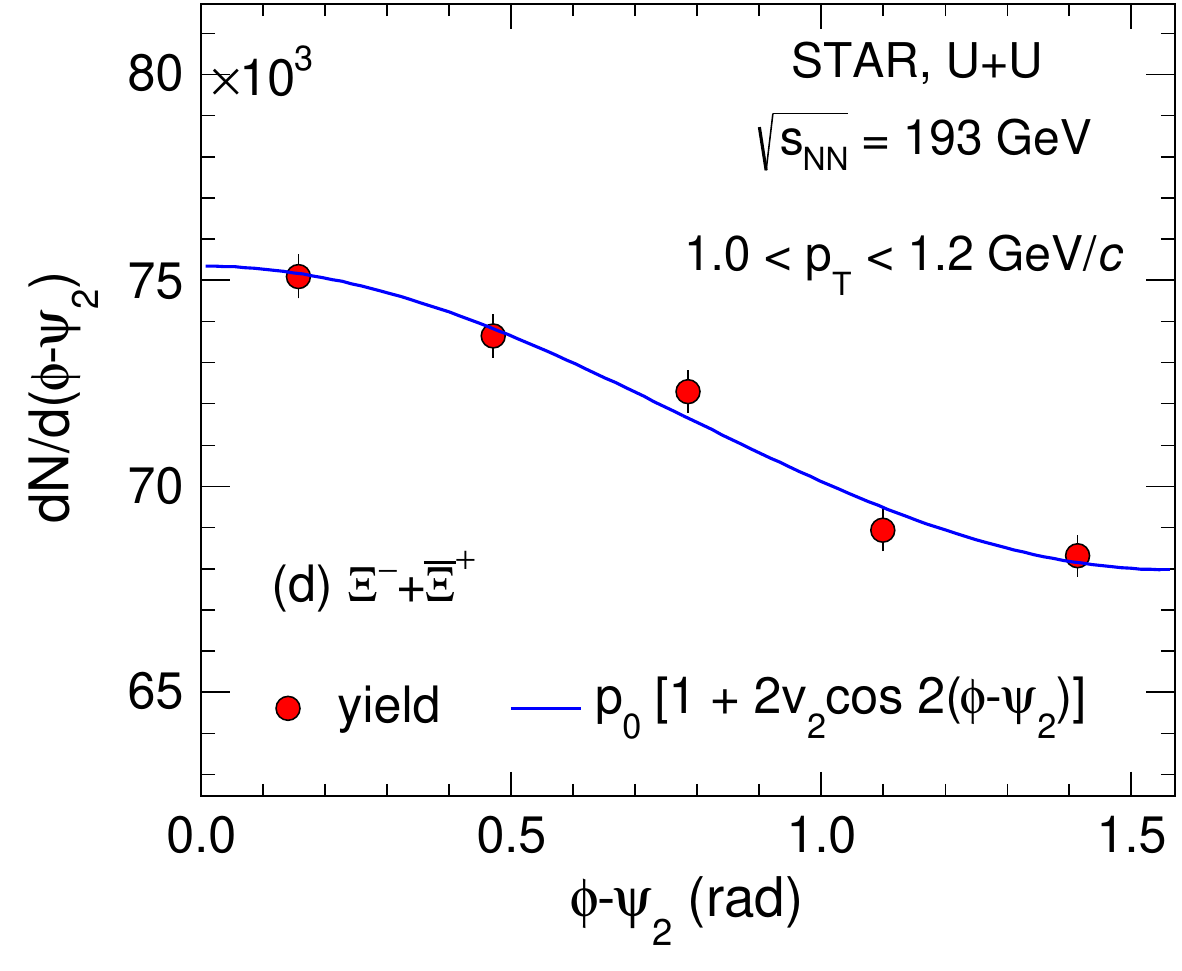}
\includegraphics[width=0.3\textwidth]{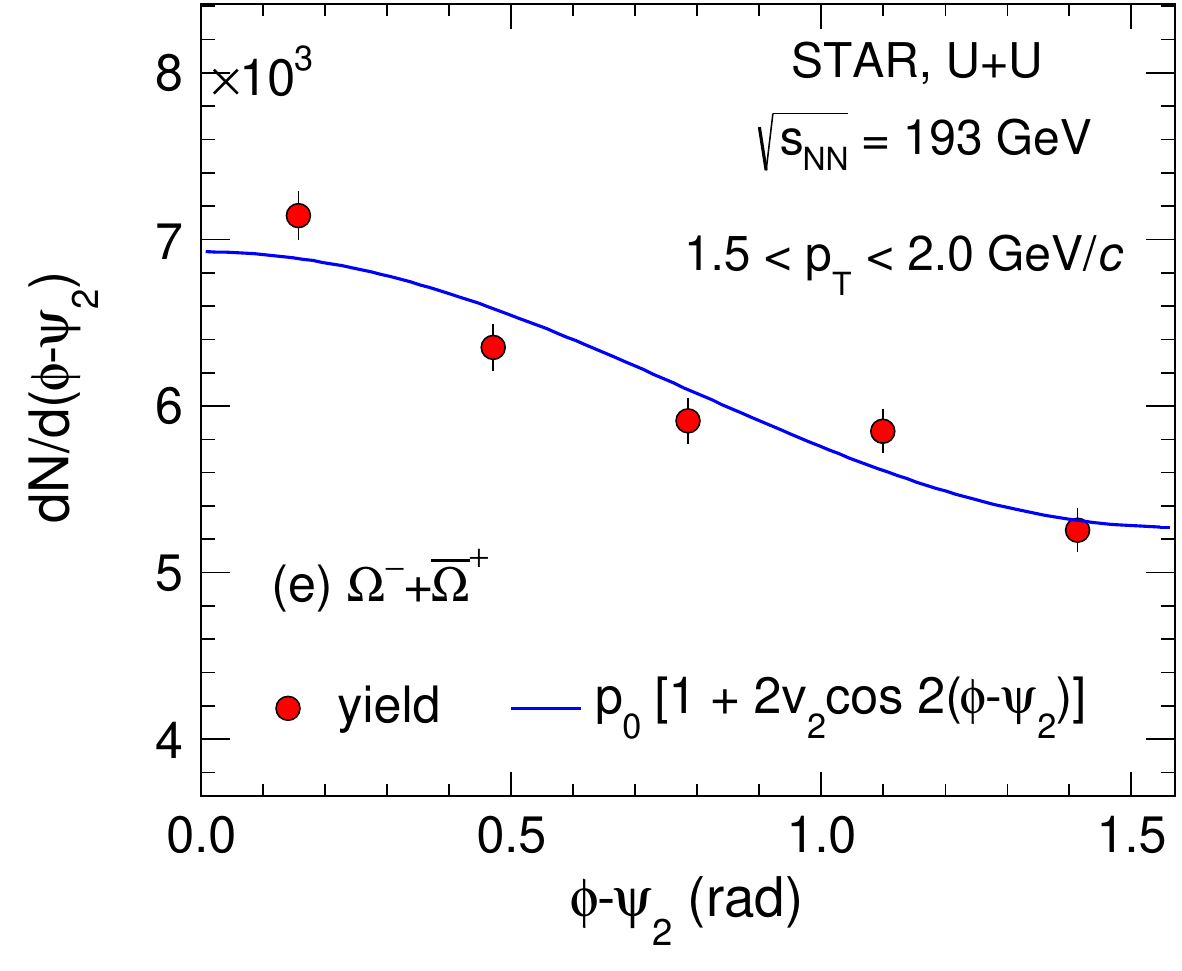}
\caption{Raw yield as a function of $\phi-\psi_{2}$ for $K^{0}_{s}$, $\phi$, $\Lambda$, $\Xi$, and $\Omega$ at mid-rapidity ($|y| <$ 1) in minimum bias U+U collisions at $\sqrt{s_{NN}}$ = 193 GeV. The solid blue line represents the fit to the data to extract $v_{2}$ for each $p_{\text{T}}$ bin. Error bars represent the statistical uncertainties. Note that in some cases errors are smaller than the markers.}
\label{fig:vnfit}
\end{figure*}

Short-lived hadrons $K^{0}_{s}$, $\phi$, $\Lambda$, $\Xi$, and $\Omega$ cannot be identified in the same way as stable hadrons $\pi$, $K$, and $p(\bar{p})$, hence their $v_{n}$ coefficients cannot directly be measured using Eq.~\ref{eq:vncorr}. Therefore, for these particles, first the raw yield of particle candidates is measured as a function of their invariant mass, transverse momentum ($p_{\text{T}}$) and azimuthal angle with respect to the event-plane angle ($\phi-\psi_{n}$). Then the yield of particles is obtained as a function of angle $\phi-\psi_{n}$ in various $p_{\text{T}}$ intervals for each centrality class. The extraction of $\phi$-meson yield is carried out by fitting the invariant mass distribution with a Breit-Wigner function plus a $2^{nd}$-order polynomial function~\cite{phimeson1}. For weak-decay particles, $K_{s}^{0}$, $\Lambda$, $\Xi$, and $\Omega$, the raw-yield is extracted using a bin-counting method~\cite{ksl1,ksl2}. For $K_{s}^{0}$ and $\Lambda$, the invariant mass region chosen for bin-counting is $\pm$ 20 MeV around their rest mass values taken from the PDG~\cite{pdg1}, which are $497.611\pm0.013$ MeV/$c^{2}$ and $1115.683\pm0.006$ MeV/$c^{2}$, respectively. For $\Xi$ and $\Omega$, the default mass window for bin counting is $\pm 10$ MeV around their rest masses from the PDG, which are $1321.71\pm0.07$ MeV/$c^{2}$ and $1672.45\pm0.29$ MeV/$c^{2}$, respectively. We observe a typical value of signal to background ratio (S/B), averaged over $p_{\text{T}}$ and $\phi-\psi_{n}$ bins, of 0.04 for $\phi$-meson, 9.18 for $K_{s}^{0}$, 1.25 for $\Lambda$, 1.85 for $\Xi$, and 0.53 for $\Omega$ in minimum bias U+U collisions.

Figure~\ref{fig:vnfit} shows examples of the particle yields as a function of $\phi-\psi_{2}$ for given $p_{\text{T}}$ ranges. The observed $v_{2}$ is obtained by fitting the yields with the functional form given by the equation 
\begin{equation}
\frac{dN}{d(\phi-\psi_{n})} = A \left( 1+ 2 \sum_{n} v_{n} \cos n(\phi-\psi_{n}) \right),
\label{eq:vnfunction}
\end{equation}
where A is a normalization parameter. Finally, the true $v_{2}$ is obtained by dividing the observed $v_{2}$ with the corresponding event plane resolution. The $p_{\text{T}}$ dependence of the flow coefficients are studied by repeating the above procedure for fixed ranges of $p_{\text{T}}$. The same procedure is used to extract higher order harmonics $v_{3}$ and $v_{4}$ with respect to $\psi_{3}$ and $\psi_{4}$.

\section{Systematic uncertainties}
\label{sys_error}

\begin{table*}[!htbp]
\centering
\caption{Systematic uncertainties on flow coefficients of $K^{0}_{s}$, $\phi$, $\Lambda$, $\Xi$, and $\Omega$ due to various sources in U+U collisions at $\sqrt{s_{NN}}$ = 193 GeV. All numbers represent percent uncertainties.}
\label{tab:syserr}
\begin{tabular}{|c|ccc|ccc|ccc|ccc|ccc|}
\hline \rule{0pt}{14pt}
Particle				& \multicolumn{3}{c|}{$K_{s}^{0}$}   
						& \multicolumn{3}{c|}{$\phi$}
						& \multicolumn{3}{c|}{$\Lambda$}
						& \multicolumn{3}{c|}{$\Xi$}
						& \multicolumn{3}{c|}{$\Omega$}			 \\[0.5ex]
\hline \rule{0pt}{14pt}
Flow order				& {$v_{2}$}  	& {$v_{3}$}  & {$v_{4}$} 	
						& {$v_{2}$}  	& {$v_{3}$}  & {$v_{4}$}
						& {$v_{2}$}  	& {$v_{3}$}  & {$v_{4}$}
						& {$v_{2}$}  	& {$v_{3}$}  & {$v_{4}$}
						& {$v_{2}$}  	& {$v_{3}$}  & {$v_{4}$}	\\[0.5ex] 
\hline \rule{0pt}{14pt}
Event Cuts				& 3				& 4			& 5	
						& 5				& 7			& 7	
						& 2				& 3			& 5
						& 4				& 4			& 6
						& 5				& 6			& 8	\\[0.5ex]
						
Track Cuts				& 2				& 7			& 11	
						& 7				& 8			& 12
						& 1				& 7			& 11
						& 5				& 9			& 12
						& 8				& 15		& 15	\\[0.5ex]	
											
PID Cuts				& 3				& 4			& 6	
						& 6				& 10		& 10	
						& 1				& 3			& 9
						& 2				& 5			& 5
						& 3				& 10		& 15	\\[0.5ex]
							
V0 Cuts					& 4				& 6			& 6	
						& --			& --		& --	
						& 1				& 6			& 5
						& 3				& 7			& 8
						& 2				& 6			& 10	\\[0.5ex]
						
Background		 		& 2				& 5			& 5	
						& 8				& 7			& 8
						& 3				& 6			& 6
						& 3				& 3			& 9
						& 3				& 8			& 10	\\[0.5ex]	
\hline \rule{0pt}{14pt}
Total			 		& 6				& 12		& 16	
						& 13			& 16		& 19
						& 4				& 12		& 17
						& 8				& 13		& 19
						& 11			& 21		& 27	\\[0.5ex]
\hline
\end{tabular}
\end{table*}
Point-by-point systematic uncertainties on the flow coefficient $v_{n}(p_{\text{T}})$ measurements are estimated by varying event selection criteria, track selection criteria, particle identification criteria, and $V^{0}$ topology criteria from their default values. The selection criterion for $z$-coordinate of the primary vertex ($V_{z}$) is varied to $\pm$ 20 cm and $\pm$ 25 cm from the default value ($\pm$ 30 cm). The DCA of the primary tracks is varied between 1.5 cm and 2.5 cm. The number of fit points is varied from 18 to 24. The $\eta$ gap for event-plane angle calculation is varied between 0.05 and 0.15. The PID selection cuts for a given particle, $|n_{\sigma}|$, is varied from $|n_{\sigma}| <$ 1.5 to $|n_{\sigma}| <$ 3.0. For weak decay particles, various topology cuts such as daughter particle DCA, $V^{0}$ DCA to primary vertex, decay length, and mass width are varied. Most of the cuts were varied $\sim$20\% from their default values. The selection criteria are each varied one at a time while keeping others at the default values. In addition to these, systematic uncertainties from combinatorial background and residual background are also estimated. The uncertainty due to the combinatorial background is estimated by using different background methods mentioned in section~\ref{comb_back}. In order to estimate uncertainty due to the residual background shape, we have used $1^{st}$ and $2^{nd}$-order polynomial functions to fit residual background. Table~\ref{tab:syserr} shows the systematic uncertainties from different sources on $v_{2}$, $v_{3}$, and $v_{4}$ for each particle. Total systematic uncertainty is calculated by adding uncertainties from different sources in quadrature. The systematic uncertainties vary with the $p_{\text{T}}$ and centrality. In general, at low $p_{\text{T}}$, they are smaller than at higher $p_{\text{T}}$ for a given centrality. 

\section{Results and Discussion}
\label{result}
In this section, the $p_{\text{T}}$ dependence of flow coefficients $v_{2}$, $v_{3}$, and $v_{4}$ is presented for strange and multi-strange hadrons at mid-rapidity ($|y| <$ 1) for minimum bias and various centrality classes in U+U collisions at $\sqrt{s_{NN}}$ = 193 GeV. 
\subsection{\texorpdfstring{$p_{\text{T}}$}{Lg} dependence of flow coefficients}
\label{ptdep}
\begin{figure*}[!htbp]
\centering
\includegraphics[width=0.7\textwidth]{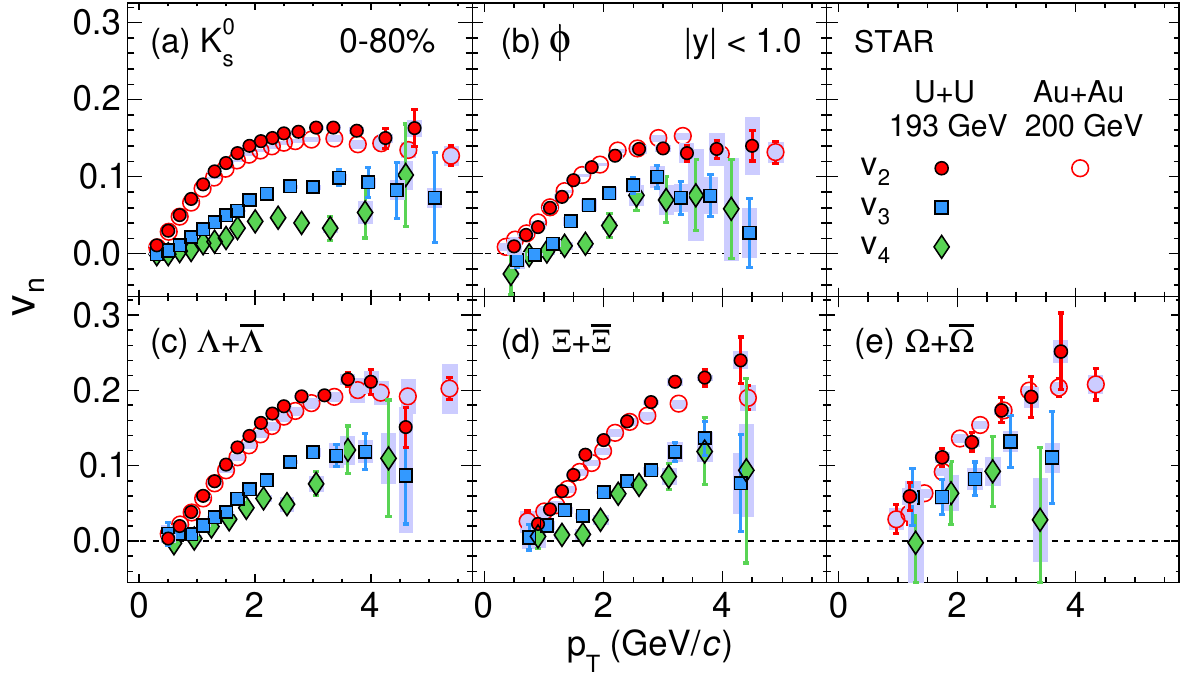}
\caption{The $p_{\text{T}}$ dependence of $v_{2}$, $v_{3}$, and $v_{4}$ for (a) $K^{0}_{s}$, (b) $\phi$, (c) $\Lambda$, (d) $\Xi$, and (e) $\Omega$ at mid-rapidity ($|y| <$ 1) in minimum bias U+U collisions at $\sqrt{s_{NN}}$ = 193 GeV. The error bars represent statistical uncertainties. The bands represent point-by-point systematic uncertainties. For comparison, published results for $v_{2}$ from Au+Au collisions at $\sqrt{s_{NN}}$ = 200 GeV are shown by open markers~\citep{mhadflow2,idflowe1}.}
\label{fig:vn_mb}
\end{figure*}

Figure~\ref{fig:vn_mb} shows the transverse momentum dependence of flow coefficients $v_{2}$, $v_{3}$, and $v_{4}$ for (a) $K^{0}_{s}$, (b) $\phi$, (c) $\Lambda$, (d) $\Xi$, and (e) $\Omega$ at mid-rapidity ($|y| <$ 1) in minimum bias U+U collisions. The flow coefficients first increase with increasing $p_{\text{T}}$ and then saturate for the intermediate $p_{\text{T}}$ region. The $p_{\text{T}}$ dependence of elliptic flow $v_{2}$ in U+U collisions is similar to that observed in Au+Au collisions at $\sqrt{s_{NN}}$ = 200 GeV. In addition, the flow coefficients show a monotonic increase with increasing $p_{\text{T}}$ reaching a maximum value at $p_{\text{T}}$ between 2-3 GeV/$c$. This maximum has a dependence on particle mass as it takes place at comparatively higher $p_{\text{T}}$ for heavier particles than for lighter particles. 

We observe that the magnitude of $v_{2}$ is greater than $v_{3}$ and $v_{4}$ in minimum bias U+U collisions for the measured $p_{\text{T}}$ range, while $v_{3}$ is comparable to $v_{4}$ for higher $p_{\text{T}}$. The non-zero values of higher-order flow coefficients (especially, $v_{3}$) for the measured $p_{\text{T}}$ range is an indication of event-by-event fluctuations in the initial energy density profile~\cite{flowm3}.   

\subsection{Centrality dependence of flow coefficients}
\begin{figure*}[!htbp]
\centering
\includegraphics[width=.98\textwidth]{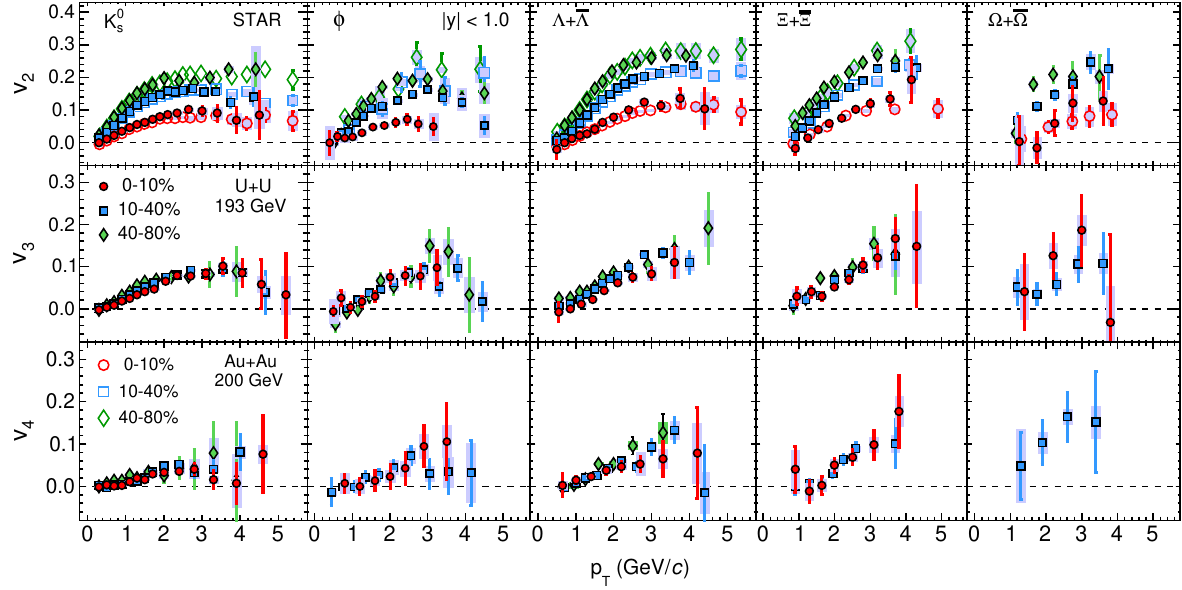}
\caption{The flow coefficients $v_{2}$, $v_{3}$, and $v_{4}$ as a function of $p_{\text{T}}$ for $K^{0}_{s}$, $\phi$, $\Lambda$, $\Xi$, and $\Omega$ at mid-rapidity ($|y| <$ 1) in U+U collisions at $\sqrt{s_{NN}}$ = 193 GeV for centrality classes 0-10\%, 10-40\%, and 40-80\%. The error bars represent statistical uncertainties. The bands represent point-by-point systematic uncertainties. For comparison, published results for $v_{2}$ from Au+Au collisions at $\sqrt{s_{NN}}$ = 200 GeV are shown by open markers~\citep{mhadflow2,idflowe1}.}
\label{fig:vn_cen}
\end{figure*}

Figure~\ref{fig:vn_cen} shows the flow coefficients of $K^{0}_{s}$, $\phi$, $\Lambda$, $\Xi$, and $\Omega$ in U+U collisions for various centrality classes. The top panels show the $p_{\text{T}}$ dependence of $v_{2}$ for these centralities. The magnitude of $v_{2}$ increases strongly from central to peripheral collisions for all particle species. The centrality dependence of $v_{2}$ is similar to the published results of $v_{2}$ in Au+Au collisions at $\sqrt{s_{NN}}$ = 200 GeV~\cite{mhadflow2,idflowe1}. This centrality dependence is expected as the eccentricity of the initial overlap region of the colliding nuclei increases from central to peripheral collisions. This observation is consistent with the interpretation from the hydrodynamic model which predicts that final state momentum anisotropy is driven by the initial spatial anisotropy~\cite{flowCentDep1}. We also observe negative values of $v_{2}$ for $\Lambda$, $\Xi$, and $\Omega$ at very low $p_{\text{T}}$ in central collisions, which suggests the strong expansion observed in hadron $p_{\text{T}}$-spectra.

The middle and bottom panels of Fig.~\ref{fig:vn_cen} present centrality dependence of $v_{3}$ and $v_{4}$ for different particle species. The $v_{3}$ measurements are carried out up to mid-central collisions for $\Omega$ due to limited statistics. For the same reason, $v_{4}$ measurements for particles except $K^{0}_{s}$ and $\Lambda$ are also carried out only for central and/or mid-central collisions. We do not observe a clear centrality dependence of $v_{3}$ and $v_{4}$. While the centrality dependence of $v_{2}$ is consistent with the scenario of hydrodynamical evolution driven by the initial participant geometrical profile shape, the lack of centrality dependence of $v_{3}$ and $v_{4}$ presumably reflects that event-by-event fluctuations are the dominant source of triangular and quadrangular shape variations rather than the shape of the collision overlap region that dominates for $v_{2}$.

\subsection{Particle mass dependence and NCQ scaling}
\begin{figure*}[!htbp]
\centering
\includegraphics[width=0.8\textwidth]{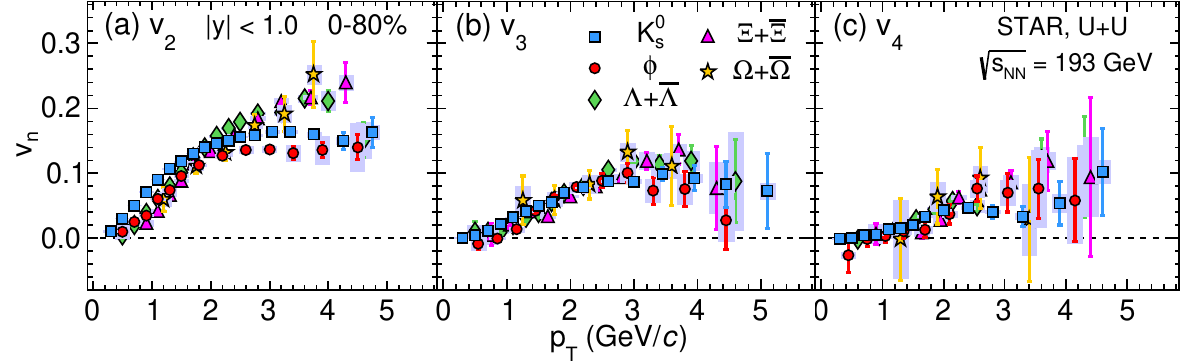}
\includegraphics[width=0.8\textwidth]{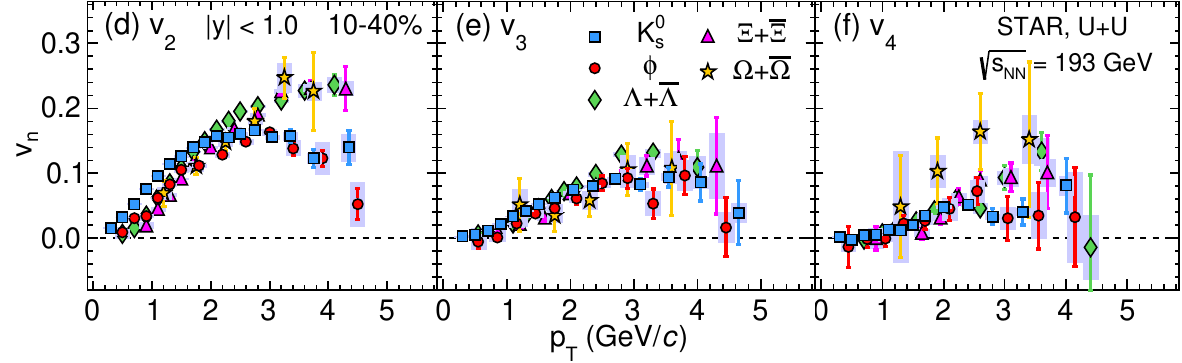}
\caption{Flow coefficients $v_{2}$, $v_{3}$, and $v_{4}$ as a function of $p_{\text{T}}$ for various particles at mid-rapidity ($|y| <$ 1), grouped together in a single panel in U+U collisions at $\sqrt{s_{NN}}$ = 193 GeV. Top panels represent $v_{n}(p_{\text{T}})$ for minimum bias (0-80\%) and bottom panels for centrality class (10-40\%). The error bars represent statistical uncertainties. The bands represent point-by-point systematic uncertainties.}
\label{fig:vn_partm}
\end{figure*}

Figure~\ref{fig:vn_partm} presents mass ordering and particle type dependence of flow coefficients $v_{n}(p_{\text{T}})$ for strange and multi-strange hadrons in minimum bias (top panels) and 10-40\% mid-central (bottom panels) U+U collisions. A clear mass ordering of elliptic flow $v_{2}$ is observed for $p_{\text{T}} <$ 2-3 GeV/$c$. In this $p_{\text{T}}$ region, the lighter mass particles have a larger $v_{2}$ than the heavier particles at a given value of $p_{\text{T}}$. This mass ordering at low $p_{\text{T}}$ can be attributed to a velocity field (i.e., radial flow) suggested by the hydrodynamical models in Refs.~\cite{flowt3,radflow1}. We observe a particle type dependence (baryon/meson) of $v_{2}$, i.e., $v_{2}^{B} > v_{2}^{M}$ in the intermediate $p_{\text{T}}$ region. The hadron type-dependence in the intermediate $p_{\text{T}}$ region has been explained by hadronization via quark coalescence and development of collective flow in the partonic phase~\cite{ncqt1,ncqt2}. The proposed mechanism for mass ordering and particle type dependence can effectively be tested by the $\phi$-meson $v_{2}$. The $\phi$ is a meson ($s\bar{s}$) and its mass is close to the $\Lambda$-baryon. Figure~\ref{fig:vn_partm} shows that the $\phi$-meson $v_{2}$ follows the $\Lambda$-baryon $v_{2}$ at low $p_{\text{T}}$, but follows the $K_{s}^{0}$-meson $v_{2}$ at intermediate $p_{\text{T}}$ for all centrality classes. The observed mass and hadron-type dependence of elliptic flow coefficients in U+U collisions is similar to those observed in 200 GeV Au+Au collisions at RHIC~\cite{mhadflow2,idflowe1}.

Figure~\ref{fig:vn_partm} also presents mass and particle type dependence of $v_{3}$ and $v_{4}$. The higher-order flow coefficients seem to show the same mass ordering at low $p_{\text{T}} <$ 2-3 GeV/$c$. However, statistical limitations make it difficult to reach a definitive conclusion on the particle type dependence at intermediate $p_{\text{T}}$ in the current analysis.

The above observation of mass and hadron-type dependence motivates us to test the number of constituent quark scaling of the flow coefficients in U+U collisions. This scaling was first observed at RHIC~\cite{mhadpro1,mhadflow1,mhadflow2,ncqe1,ncqe2}, where it was suggested that if $v_{n}$ of identified hadrons are scaled by the number of constituent quarks ($n_{q}$) and evaluated as a function of transverse kinetic energy per constituent quark number ($KE_{\text{T}}/n_{q}$), then the scaled values for all particle species will have an approximate similar magnitude and dependence on $KE_{\text{T}}/n_{q}$. The transverse kinetic energy is defined as $KE_{\text{T}} = m_{T} - m_{0}$, where $m_{T} = \sqrt{p_{\text{T}}^{2} + m_{0}^{2}}$ and $m_{0}$ is the rest mass of the hadron. This scaling is known as the number of constituent quark (NCQ) scaling. 
\begin{figure*}[!htbp]
\centering
\includegraphics[width=0.54\textwidth]{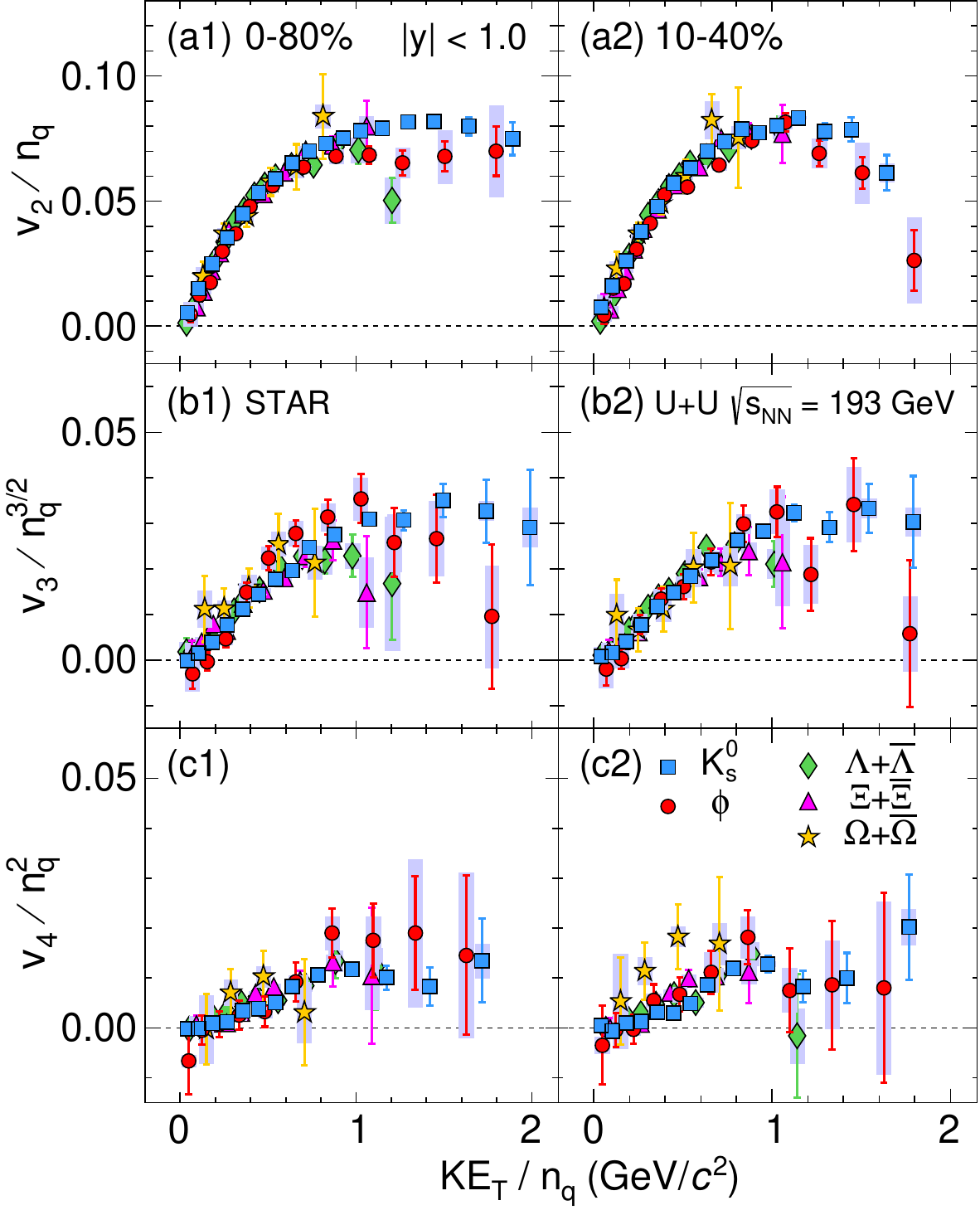}
\caption{Flow coefficients $v_{2}$, $v_{3}$, and $v_{4}$ as a function of transverse kinetic energy $KE_{\text{T}}/n_{q}$ for various particles at mid-rapidity ($|y| <$ 1) in U+U collisions at $\sqrt{s_{NN}}$ = 193 GeV, scaled by the number of constituent quarks $(n_{q})$ to the power $n/2$. Left panels represent results for minimum bias (0-80\%) and right panels for centrality class (10-40\%). The error bars represent statistical uncertainties. The bands represent point-by-point systematic uncertainties.}
\label{fig:vn_modncq}
\end{figure*}

Figure~\ref{fig:vn_modncq} shows the results of $v_{n}$ coefficients scaled by $n_{q}^{n/2}$ as a function of $KE_{\text{T}}/n_{q}$, for strange and multi-strange hadrons in U+U collisions at $\sqrt{s_{NN}}$ = 193 GeV. We observe that the NCQ scaling for current measurements holds within experimental uncertainties for each harmonic order $n$. The values of $v_{n}/n_{q}^{n/2}$ as a function of $KE_{\text{T}}/n_{q}$ lie on a single curve for all the particle species within a $\pm$15\% range. The observed NCQ scaling of $v_{n}$ coefficients in experimental data indicates the development of partonic collectivity during the QGP phase in heavy-ion collisions. Such a scaling of identified hadrons also suggests the formation of hadrons through quark coalescence or parton recombination in the intermediate $p_{\text{T}}$ range (2.0 GeV/$c$ $< p_{\text{T}} <$ 4.0 GeV/$c$)~\cite{ncqt1,ncqt2}. Although there are large differences in the collision geometry between U+U and Au+Au collisions, the hydrodynamical evolution and the coalescence mechanism for hadron formation remain key features of QGP drops created in nucleus-nucleus collisions.

\subsection{Eccentricity scaling of \texorpdfstring{$v_{n}$}{Lg} coefficients}
\begin{figure*}[!htbp]
\centering
\includegraphics[width=0.8\textwidth]{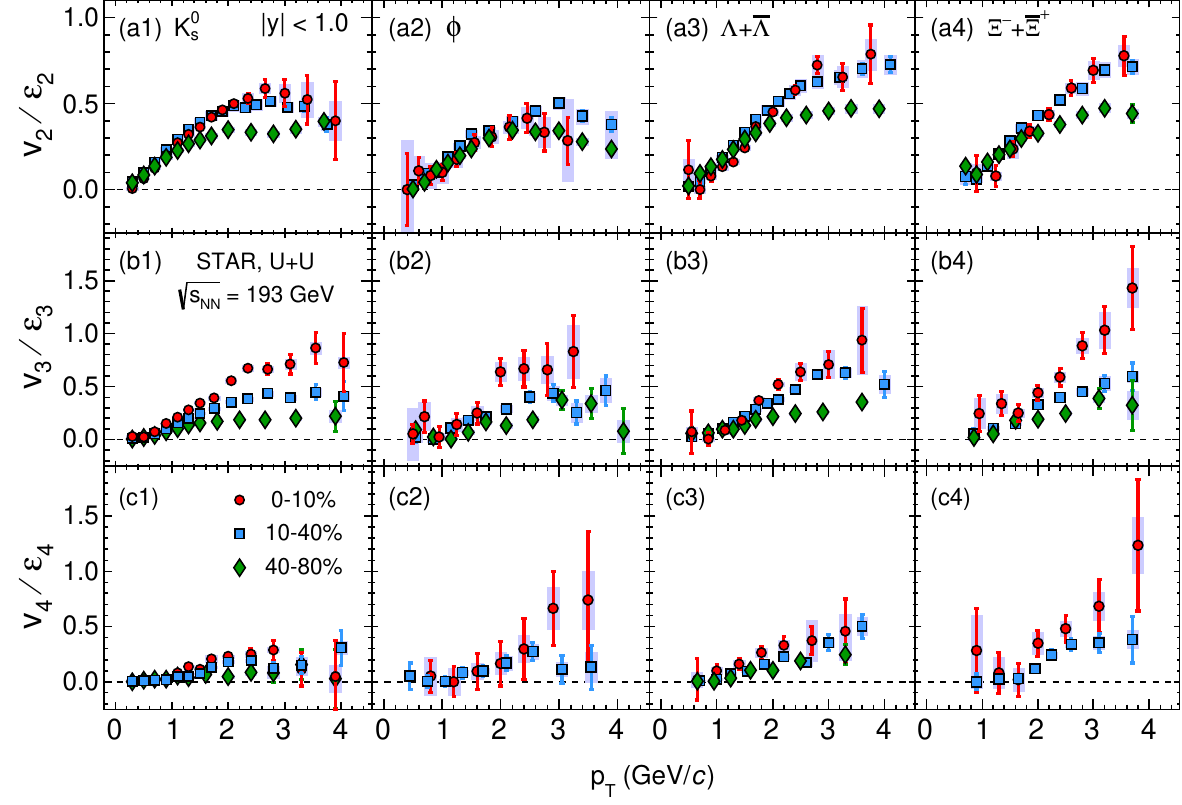}
\caption{The flow coefficients $v_{n}$ scaled by $\varepsilon_{n}\left\lbrace 2\right\rbrace$ as a function of $p_{\text{T}}$ for $K^{0}_{s}$, $\phi$, $\Lambda$, and $\Xi$ at mid-rapidity ($|y| <$ 1) in U+U collisions at $\sqrt{s_{NN}}$ = 193 GeV for centrality intervals 0-10\%, 10-40\%, and 40-80\%. The error bars represent statistical uncertainties. The bands represent point-by-point systematic uncertainties.}
\label{fig:vnen}
\end{figure*}
\begin{figure*}[!htbp]
\centering
\includegraphics[width=0.8\textwidth]{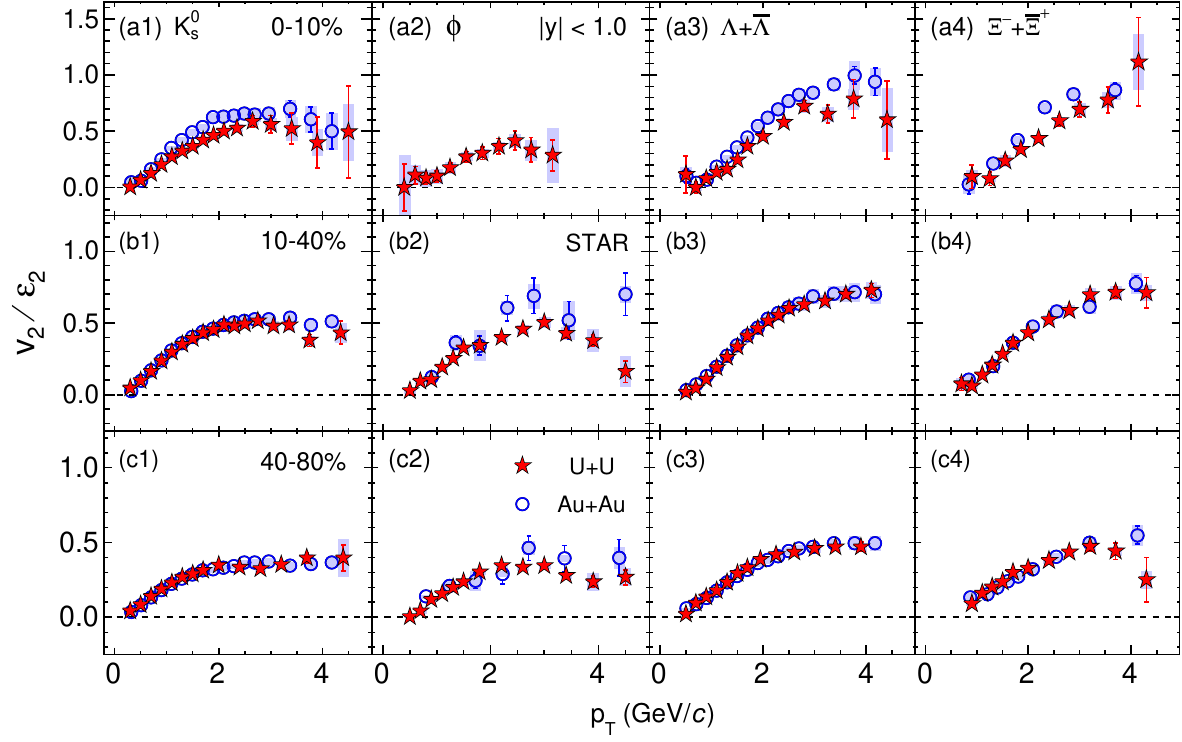}
\caption{Elliptic flow $v_{2}$ scaled by $\varepsilon_{n}\left\lbrace 2\right\rbrace$ as a function of $p_{\text{T}}$ for $K^{0}_{s}$, $\phi$, $\Lambda$, and $\Xi$ at mid-rapidity ($|y| <$ 1) in U+U collisions at $\sqrt{s_{NN}}$ = 193 GeV for centrality intervals 0-10\%, 10-40\%, and 40-80\%. The error bars represent statistical uncertainties. The bands represent point-by-point systematic uncertainties. For comparison, published results for $v_{2}/\varepsilon_{2}\left\lbrace part\right\rbrace$ from Au+Au collisions at $\sqrt{s_{NN}}$ = 200 GeV are shown by open circles~\citep{mhadflow2,idflowe1}.}
\label{fig:v2e2}
\end{figure*}

Uranium nuclei have an intrinsic prolate shape, which results in various initial state collision configurations~\cite{uu3,uu4}. Even in fully overlapping U+U collisions, owing to the deformation, the initial overlap zone can give rise to different initial spatial anisotropies compared to Au+Au collisions. In this section, we present $v_{n}$ coefficients scaled by the initial spatial eccentricity $\varepsilon_{n}$ to explore the dependence of final state momentum space anisotropy on the initial collision geometry in heavy-ion collisions. This will provide insight into the underlying dynamics driving the shape and size dependence of the collectivity developed in the heavy-ion collisions at RHIC.
 
In Fig.~\ref{fig:vnen}, we show the ratio $v_{n}/\varepsilon_{n}$ for various particles in 0-10\%, 10-40\%, and 40-80\% centrality intervals in U+U collisions at $\sqrt{s_{NN}}$ = 193 GeV. The eccentricity-scaled $v_{2}$ values exhibit a distinct centrality and particle type dependence, and the dependence varies with $p_{\text{T}}$, presumably an indication of convolution of hydrodynamical flow and coalescence formation dynamics in heavy ion collisions. The ratios $v_{3}/\varepsilon_{3}$ and $v_{4}/\varepsilon_{4}$ follow the same general trend of lower $v_{n}/\varepsilon_{n}$ in more peripheral collisions, but $v_{4}/\varepsilon_{4}$ are not conclusive with the current statistics.

Previous STAR measurements of Au+Au collisions at 200 GeV have shown that the $v_{2}$ values scaled by the participant eccentricity are larger in central collisions compared to peripheral collisions, which is an indication that stronger collectivity develops in more central collisions~\cite{flowe3}. In Fig.~\ref{fig:v2e2}, we compare the ratio $v_{2}/\varepsilon_{2}$ with the published results from Au+Au collisions at 200 GeV. We observe that the magnitude of $v_{2}/\varepsilon_{2}$ in mid-central (10-40\%) and peripheral collisions (40–80\%) is similar in both U+U and Au+Au collisions. However, the magnitude of $v_{2}/\varepsilon_{2}$ in most central collisions (0-10\%) is higher for Au+Au collisions compared to U+U collisions. This observation is the reverse of the expectation that $v_{2}/\varepsilon_{2}$, which is a measure of collectivity, should be higher in U+U collisions~\cite{flowe3,idflowe1}. This same qualitative feature was reported in a recent publication~\cite{vnen1} that used AMPT model calculations to study collective flow in these systems. We note that there is a large difference of ellipticity in central collisions of these systems, with $\varepsilon_{2}^{\text{U}}/\varepsilon_{2}^{\text{Au}}$ $\sim$1.5. The observation that $v_{2}/\varepsilon_{2}$ is greater in central Au+Au than that in central U+U suggests that in collisions of highly deformed nuclei such as Uranium, dynamics beyond eccentricity scaling may play an important role.

\subsection{Eccentricity scaling of \texorpdfstring{$v_{n}/n_{q}$}{Lg}}
The centrality and system size dependence of elliptic flow $v_{2}$ depends on a combination of eccentricity, viscosity of the fluid and the extent of equilibrium reached in heavy ion collisions~\cite{vnen2}. The ideal hydrodynamic model predicts that $v_{2}$ scaled by the eccentricity is independent of centrality and size of the collision system. The results presented in Ref.~\cite{ncqe2} show that the charged hadron $v_{2}$ scaled by ellipticity in Au+Au and Cu+Cu collisions at $\sqrt{s_{NN}}$ = 200 GeV is independent of the collision centralities and colliding system size. However, other experimental results have shown that $v_{2}$ divided by participant eccentricity in Au+Au and Cu+Cu collisions do not show scaling amongst different collision centralities~\cite{flowe3,idflowe1}. 

\begin{figure*}[!htbp]
\centering 
\includegraphics[width=0.7\textwidth]{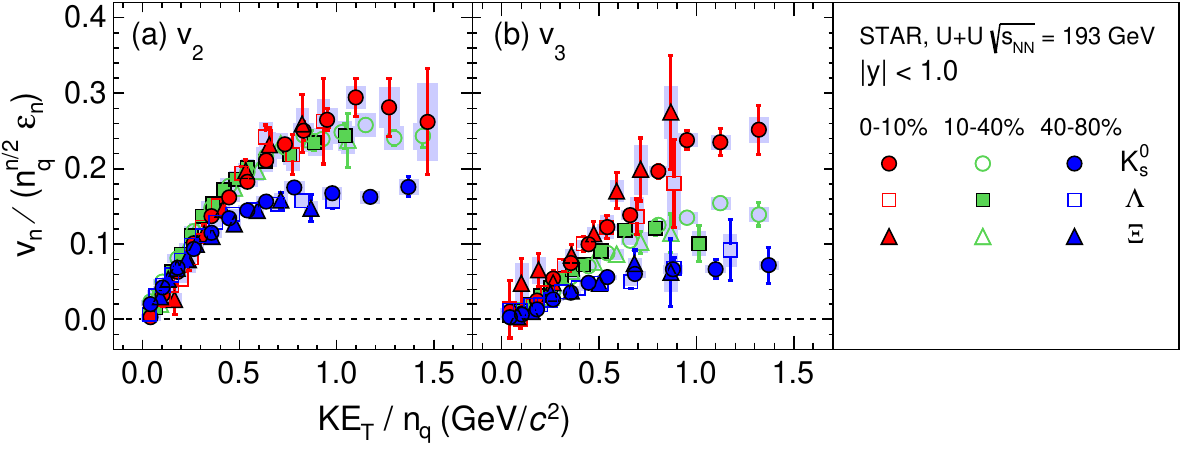}
\caption{$v_{n}$ coefficients, scaled by the number of constituent quarks $(n_{q})$ to the power $n/2$ and participant eccentricity $\varepsilon_{n}$, of identified particles versus $(m_{T}-m_{0})/n_{q}$ for three centrality bins in U+U collisions at $\sqrt{s_{NN}}$ = 193 GeV. The error bars represent statistical uncertainties. The bands represent point-by-point systematic uncertainties.}
\label{fig:vnennq}
\end{figure*}

In order to analyze the centrality dependence of the $n_{q}$-scaled flow coefficients in U+U collisions, we divide $v_{n}/n_{q}$ by the participant eccentricity $\varepsilon_{n}$. The results are depicted in Fig.~\ref{fig:vnennq}. The plots (a) and (b) show the doubly scaled quantities from three centrality bins as a function of $(m_{T}-m_{0})/n_{q}$ for $v_{2}$ and $v_{3}$, respectively. Both the plots show an initial rise and a turn over to a flat region for $(m_{T}-m_{0})/n_{q} >$ 1.0 GeV/$c^{2}$. Our measurements show that both the flow harmonics $v_{2}$ and $v_{3}$, at a given centrality, of all hadrons are scaled similar to the case of minimum bias collisions as in Fig.~\ref{fig:vn_modncq}. However, there is no scaling observed amongst different collision centralities. Therefore, the universal
scaling with eccentricity as suggested by ideal hydrodynamics is not supported by the current data-set.
 
\subsection{Model comparisons}
\begin{figure*}[!htbp]
\centering 
\includegraphics[width=0.8\textwidth]{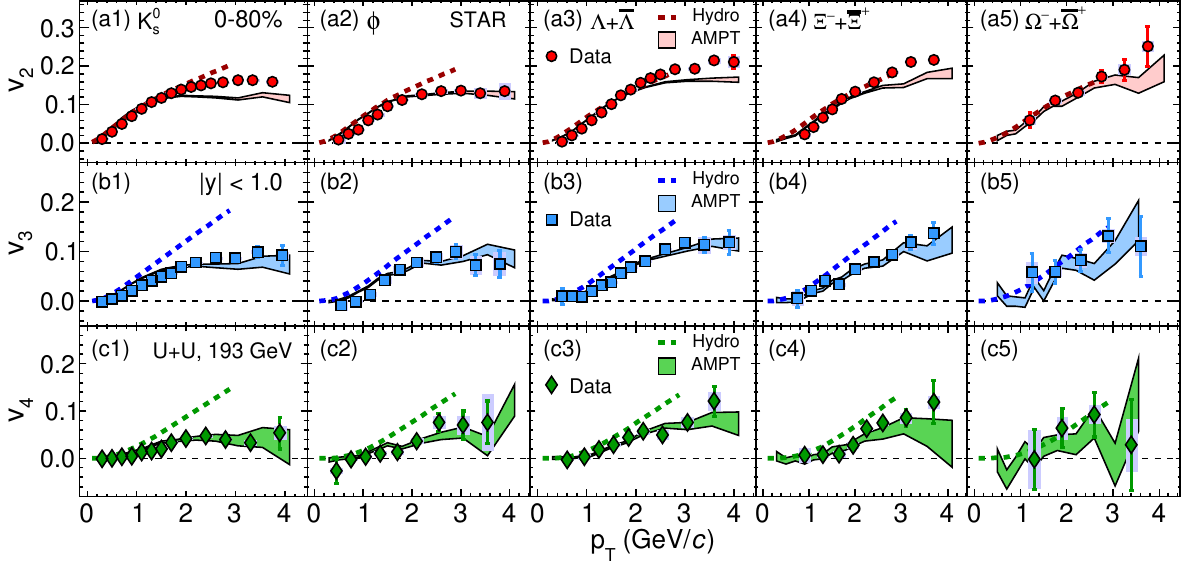}
\caption{Measured coefficients $v_{2}$, $v_{3}$, and $v_{4}$ in the top, middle, and bottom panel, respectively, as a function of $p_{\text{T}}$ for various particles at mid-rapidity ($|y| <$ 1) in minimum bias (0-80\%) U+U collisions at $\sqrt{s_{NN}}$ = 193 GeV, compared with AMPT string melting and ideal hydrodynamic model calculations. AMPT results are shown by colored bands while ideal hydrodynamic results are shown by colored dashed lines.}
\label{fig:vnmodel080}
\end{figure*}

\begin{figure*}[!htbp]
\centering 
\includegraphics[width=0.8\textwidth]{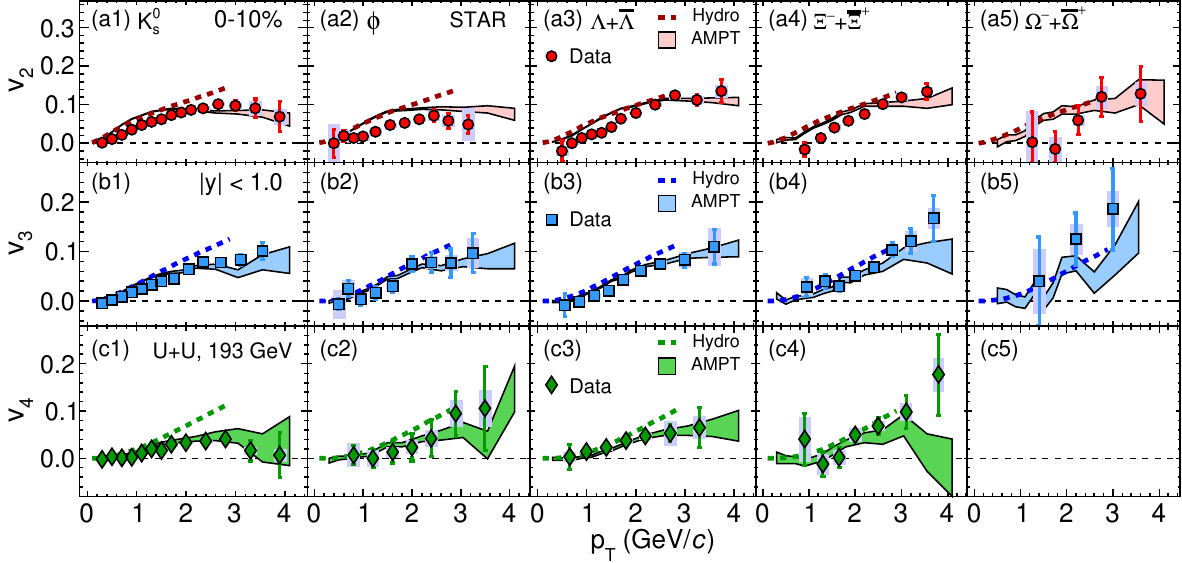}
\caption{Measured coefficients $v_{2}$, $v_{3}$, and $v_{4}$ in the top, middle, and bottom panel, respectively, as a function of $p_{\text{T}}$ for various particles at mid-rapidity ($|y| <$ 1) for centrality class 0-10\% in U+U collisions at $\sqrt{s_{NN}}$ = 193 GeV, compared with AMPT string melting and ideal hydrodynamic model calculations. AMPT results are shown by colored bands while ideal hydrodynamic results are shown by colored dashed lines.}
\label{fig:vnmodel010}
\end{figure*}

\begin{figure*}[!htbp]
\centering 
\includegraphics[width=0.8\textwidth]{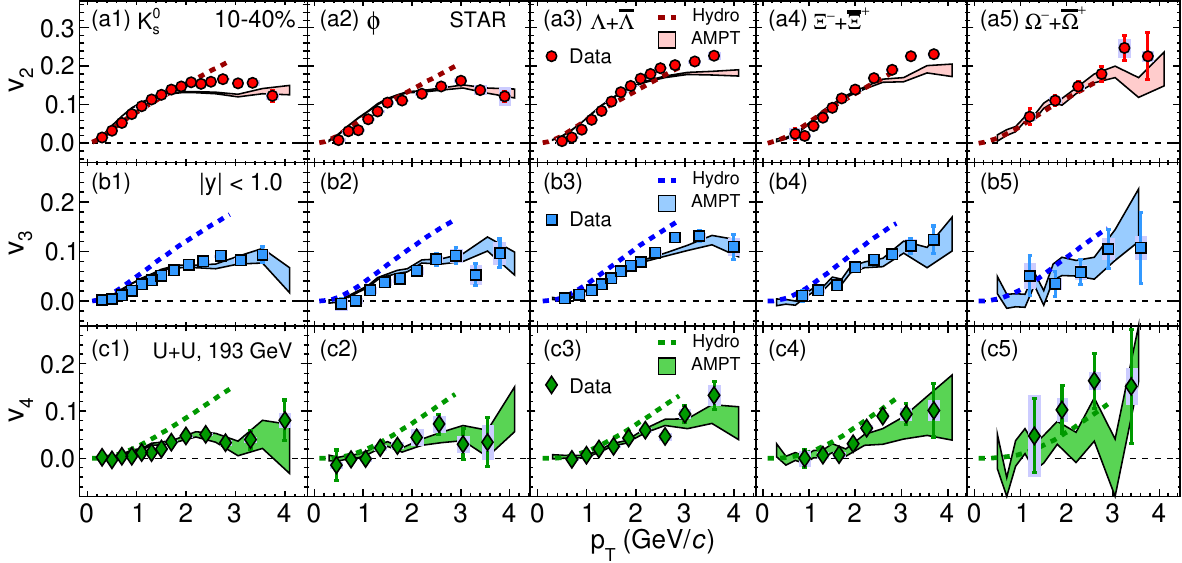}
\caption{Measured coefficients $v_{2}$, $v_{3}$, and $v_{4}$ in the top, middle, and bottom panel, respectively, as a function of $p_{\text{T}}$ for various particles at mid-rapidity ($|y| <$ 1) for centrality class 10-40\% in U+U collisions at $\sqrt{s_{NN}}$ = 193 GeV, compared with AMPT string melting and ideal hydrodynamic model calculations. AMPT results are shown by colored bands while ideal hydrodynamic results are shown by colored dashed lines.}
\label{fig:vnmodel1040}
\end{figure*}

The $v_{n}$ measurements for $K^{0}_{s}$, $\phi$, $\Lambda$, $\Xi$, and $\Omega$ in U+U collisions are compared to the hydrodynamic and transport model calculations in Figs.~\ref{fig:vnmodel080},~\ref{fig:vnmodel010}, and ~\ref{fig:vnmodel1040} for 0-80\%, 0-10\%, and 10-40\% centrality. The results from an ideal hydrodynamic model are shown by the colored dashed lines. Results from a multi-phase transport model (AMPT) are displayed by the colored bands.

The hydrodynamic model is based on the event-by-event 3+1 dimensional hydrodynamical calculations with a lattice QCD equation of state  and $\eta/s = 0$~\cite{hydro1}. The hydrodynamical calculations are able to describe the basic features of $v_{n}$ measurements at low $p_{\text{T}}$~\cite{hydro2,hydro3}. Mass ordering of $v_{n}$ coefficients are observed for strange and multi-strange hadrons in the low-$p_{\text{T}}$ region ($p_{\text{T}} <$ 2 GeV/$c$). The model is also able to predict the $p_{\text{T}}$ and centrality dependence of flow coefficients in the relatively low-$p_{\text{T}}$ region. The ideal hydrodynamic calculation deviates from data significantly at higher $p_{\text{T}}$, presumably due to viscous corrections and/or onset of different dynamics. 

In addition to the hydrodynamical calculations, $v_{n}$ measurements for $K^{0}_{s}$, $\phi$, $\Lambda$, $\Xi$, and $\Omega$ are compared to the results from the AMPT model version 2.26t9v~\cite{ampt1,ampt2,ampt3}. We have used the string melting version of the AMPT model, which incorporates both partonic and hadronic interactions. The AMPT model uses the Heavy Ion Jet Interaction (HIJING) model~\cite{hijing} for the initial conditions. The scatterings among hadrons are described by a relativistic transport (ART) model~\cite{art}. In the AMPT string melting version, hadrons are produced from the string fragmentation in the HIJING model, and are converted to their valence quarks and anti-quarks. Their evolution in space and time is modeled by the Zhang$'$s parton cascade (ZPC) model~\cite{zpc}. The input parameters such as the Lund string fragmentation parameters (a = 0.55, b = 0.15 $\rm{GeV^{-2}}$) are taken from Ref.~\cite{ampt4}. A 3 mb cross-section was used for parton-parton scattering to generate the AMPT data set, which corresponds to the parton screening mass $\mu$ = 2.2650 $\rm{fm^{-1}}$ and strong coupling constant $\alpha_{s}$ = 0.33. The AMPT model is modified to incorporate the deformation (prolate shape) of the Uranium nucleus. Various initial state configurations of deformed U+U collisions like tip-tip, body-body, and body-tip are implemented in the model. Details of the implementation and deformation parameter can be found in Refs.~\cite{uu3,uu4}. For the current analysis, a total of $\sim$5 million minimum bias U+U collisions with all possible configurations without selection of specific configurations are used. 


We observed that the AMPT string melting model with a 3 mb parton scattering cross-section, which includes hadronization via the parton coalescence mechanism, agrees well with the U+U collisions data for all flow harmonics within statistical uncertainties. It predicts mass ordering at low $p_{\text{T}}$ and a hadron type dependence in the intermediate $p_{\text{T}}$ region that are both similar to what is seen the experimental measurements. It also reproduces the transverse momentum and centrality dependence of flow coefficients in U+U collisions at $\sqrt{s_{NN}}$ = 193 GeV.
   
\begin{figure*}[!htbp]
\centering
\includegraphics[width=0.5\textwidth]{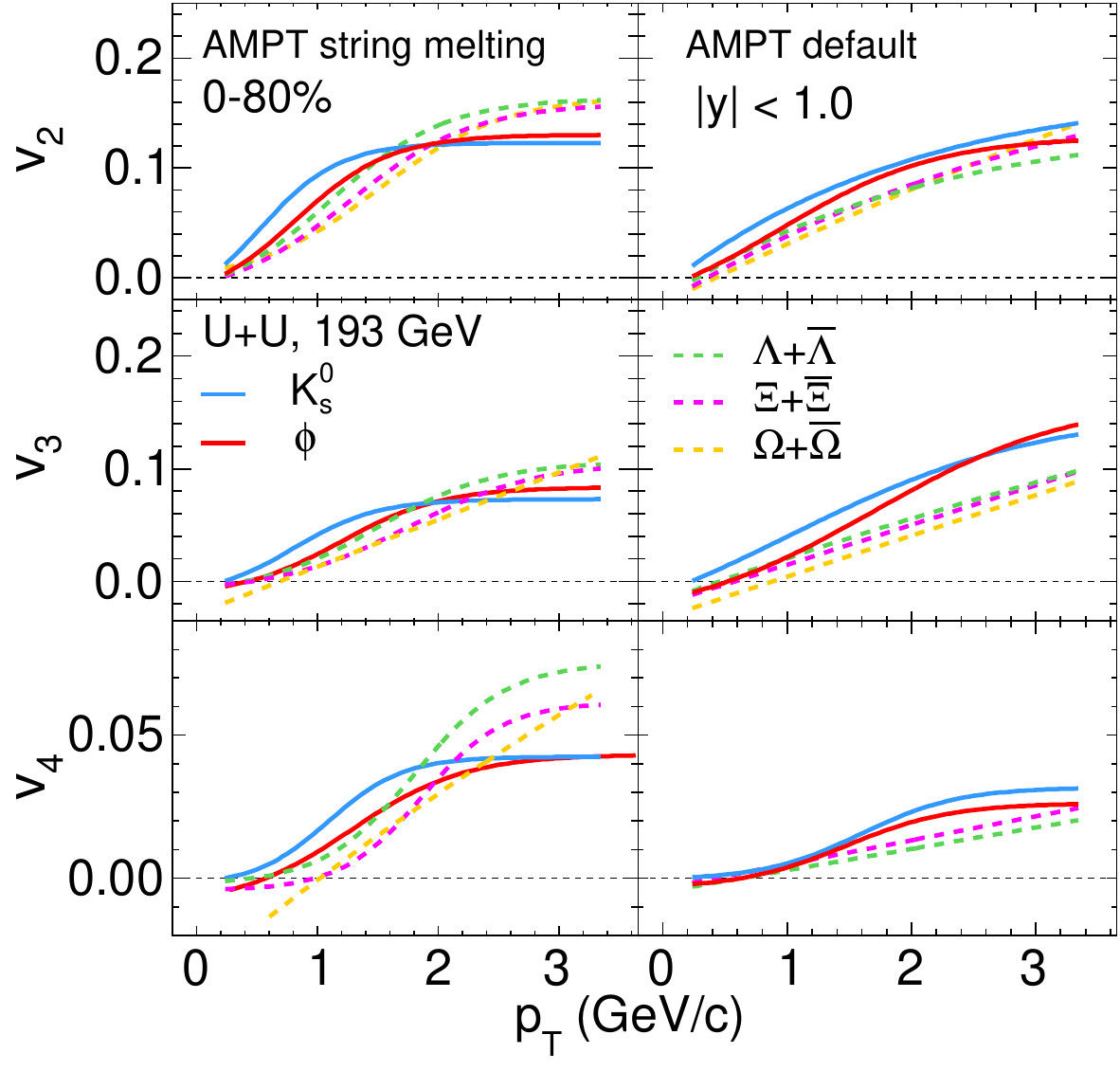}
\caption{Comparison of strange and multi-strange hadron $v_{n}(p_{\text{T}})$ at mid-rapidity ($|y| <$ 1.0) in minimum bias U+U collisions at $\sqrt{s_{NN}}$ = 193 GeV between AMPT default and string melting version. The solid lines and dashed lines represent the $v_{n}$ values for mesons and baryons, respectively.}
\label{fig:vn_amptsmdef}
\end{figure*}

Lastly, we compare $v_{n}$ measurements for $K^{0}_{s}$, $\phi$, $\Lambda$, $\Xi$, and $\Omega$ between the default and string melting version of the AMPT model. The comparison is shown in Fig.~\ref{fig:vn_amptsmdef}. Unlike the version with string melting, the AMPT default version is only able to reproduce the mass ordering in the low-$p_{\text{T}}$ region. These observations suggest that the parton degrees of freedom in the string melting scenario play an essential role leading to the particle-type dependence of $v_{n}$ coefficients at intermediate $p_{\text{T}}$.

\subsection{\texorpdfstring{$v_{n}$}{Lg} ratios}
\label{vnratio}
It has been proposed from previous measurements at RHIC~\cite{vnratio1,vnratio2} that the higher-order flow harmonics $v_{n}$ might be proportional to $v_{2}^{n/2}$, with observations showing that the ratios $v_{n}/v_{2}^{n/2}$ are independent of $p_{\text{T}}$ over the $p_{\text{T}}$ range measured. Recent measurements at the LHC~\cite{vnratio3,vnratio4,vnratio5} similarly exhibit only a weak $p_{\text{T}}$ dependence of the $v_{n}/v_{2}^{n/2}$ ratios.
\begin{figure*}[!htbp]
\centering
\includegraphics[width=0.9\textwidth]{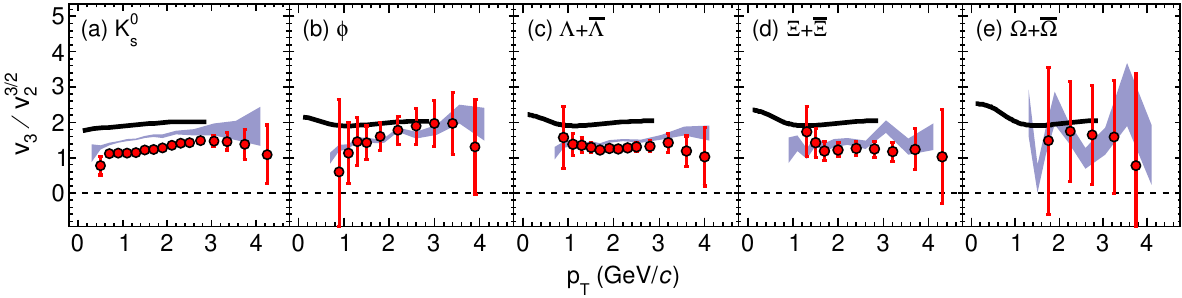}
\includegraphics[width=0.9\textwidth]{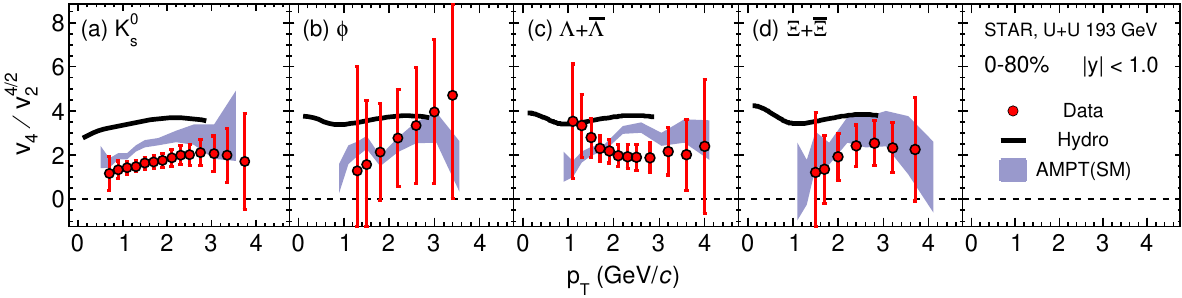}
\caption{Ratios of $v_{n}$ coefficients for $K^{0}_{s}$, $\phi$, $\Lambda$, $\Xi$, and $\Omega$ at mid-rapidity ($|y| <$ 1.0) in minimum bias U+U collisions at $\sqrt{s_{NN}}$ = 193 GeV compared with AMPT(SM) and ideal hydrodynamic model. Error bars represent statistical uncertainties. Results from the AMPT model are shown by the colored bands and hydrodynamic model by black solid lines.}
\label{fig:vn_ratios}
\end{figure*}

In Fig.~\ref{fig:vn_ratios}, we present $v_{n}$ ratios, $v_{3}/v_{2}^{3/2}$ and $v_{4}/v_{2}^{4/2}$ for $K^{0}_{s}$, $\phi$, $\Lambda$, $\Xi$, and $\Omega$ at mid-rapidity ($|y| <$ 1.0) in minimum bias U+U collisions at $\sqrt{s_{NN}}$ = 193 GeV. These ratios are compared with the corresponding results from the AMPT string melting and hydrodynamical model calculations. We observe a weak $p_{\text{T}}$ dependence of both of these ratios for mesons ($K^{0}_{s}$ and $\phi$) as well as for baryons ($\Lambda$, $\Xi$, and $\Omega$). Similar observations are found in both minimum bias (0-80\%) and mid-central (10-40\%) U+U collisions. The ideal hydrodynamical model results show a weak $p_{\text{T}}$ dependence of the $v_{n}$ ratios, similar to our measurement, but overestimate the magnitude of these ratios. AMPT string melting model results agree well with the data within statistical uncertainties. 

\section{SUMMARY}
\label{summary} 
In summary, we have reported measurements of the elliptic, triangular and quadrangular flow coefficients of $K^{0}_{s}$, $\phi$, $\Lambda$, $\Xi$, and $\Omega$ at mid-rapidity for minimum bias and various centrality intervals in $\sqrt{s_{NN}}$ = 193 GeV U+U collisions at RHIC. The $v_{n}$ coefficients are calculated as a function of transverse momentum with the $\eta$ sub-event plane method. An $\eta$ gap between the positive and negative pseudorapidity regions is used to reduce correlations not related to the anisotropic flow (i.e., non-flow). The magnitude of $v_{2}$ is found to be greater than $v_{3}$ and $v_{4}$ in minimum bias U+U collisions. The $v_{n}$ coefficients increase from central to peripheral collisions for all particle species in U+U collisions. This observation is in agreement with the observed centrality dependence of elliptic flow $v_{2}$ in Au+Au collisions at RHIC. The increase is more pronounced for elliptic flow $v_{2}$ compared to higher-order flow harmonics, which reflects dominance of collision geometry on the origin of elliptic flow, while higher-order flow harmonics are more susceptible to event-by-event fluctuations in the initial energy density distribution of participating nucleons. This scenario is supported by the observation of $v_{3}$ and $v_{4}$ having much smaller centrality dependence in contrast to distinct centrality dependence of elliptic flow $v_{2}$ in U+U collisions.

We observe a mass ordering of $v_{2}$ at low $p_{\text{T}} <$ 2-3 GeV/$c$ and a hadron-type dependence at intermediate $p_{\text{T}}$ for minimum bias and different centrality intervals. Higher order flow harmonics show similar trends within experimental uncertainties. The observation of a mass hierarchy of flow coefficients at low $p_{\text{T}}$ indicates a hydrodynamic expansion (radial flow) of the collision system. 

The measurements are compared with ideal hydrodynamical and transport model calculations. The model calculations predict the same mass ordering at low $p_{\text{T}}$ as in the data. The ideal hydrodynamical calculations over-predict the values of flow coefficients at higher $p_{\text{T}} >$ 2 GeV/$c$, which suggests the need for viscous correction and/or additional dynamics. The AMPT string melting model calculations describe the measurements within statistical uncertainties. Comparison between AMPT string melting and default configuration with the measurements suggests that the hadron production via the quark coalescence mechanism is responsible for the development of the mass ordering and hadron-type dependence of the anisotropic flow at RHIC.

Our measurements also exhibit constituent quark scaling of $v_{2}$ in the intermediate $p_{\text{T}}$ region for strange as well as multi-strange hadrons, which are expected to have small hadronic interaction cross-sections. The sizable $v_{2}$ values for multi-strange hadrons indicate collectivity of the medium produced in U+U collisions at RHIC. We also observe negative values of $v_{2}$ for $\Lambda$, $\Xi$, and $\Omega$ at very low $p_{\text{T}}$ in central U+U collisions, which shows the strong expansion observed in hadron $p_{\text{T}}$-spectra analysis. The higher-order harmonics show a modified NCQ scaling, i.e. $v_{n}$ scaled by $n_{q}^{n/2}$ follows a common trend for all particles as a function of $KE_{\text{T}}/n_{q}$. 
 
We find that the ratio $v_{2}/\varepsilon_{2}$ is higher in more central collisions compared to peripheral collisions, especially at intermediate $p_{\text{T}}$ in U+U collisions at $\sqrt{s_{NN}}$ = 193 GeV. $v_{3}/\varepsilon_{3}$ and $v_{4}/\varepsilon_{4}$ follow the same general trend. We have compared $v_{2}/\varepsilon_{2}$ in U+U collisions with the published results from Au+Au collisions at $\sqrt{s_{NN}}$ = 200 GeV, and found that the ratio in more central collisions is higher for Au+Au collisions than that for U+U collisions. This could be due to the deformed shape of the Uranium nucleus.

We observed a weak $p_{\text{T}}$ dependence of the $v_{n}$ ratios $v_{3}/v_{2}^{3/2}$ and $v_{4}/v_{2}^{4/2}$ for mesons and baryons. The ideal hydrodynamical model results also show similar weak $p_{\text{T}}$ dependence of the $v_{n}$ ratios but over-estimate the magnitude of these ratios. AMPT string melting model results agree well with the data within statistical uncertainties.

\section{ACKNOWLEDGEMENTS}
We thank the RHIC Operations Group and RCF at BNL, the NERSC Center at LBNL, and the Open Science Grid consortium for providing resources and support. This work was supported in part by the Office of Nuclear Physics within the U.S. DOE Office of Science, the U.S. National Science Foundation, the Ministry of Education and Science of the Russian Federation, National Natural Science Foundation of China, Chinese Academy of Science, the Ministry of Science and Technology of China and the Chinese Ministry of Education, the Higher Education Sprout Project by Ministry of Education at NCKU, the National Research Foundation of Korea, Czech Science Foundation and Ministry of Education, Youth and Sports of the Czech Republic, Hungarian National Research, Development and Innovation Office, New National Excellency Programme of the Hungarian Ministry of Human Capacities, Department of Atomic Energy and Department of Science and Technology of the Government of India, the National Science Centre of Poland, the Ministry of Science, Education and Sports of the Republic of Croatia, RosAtom of Russia and German Bundesministerium f\"ur Bildung, Wissenschaft, Forschung and Technologie (BMBF), Helmholtz Association, Ministry of Education, Culture, Sports, Science, and Technology (MEXT) and Japan Society for the Promotion of Science (JSPS). We also thank Dr. Victor Roy for providing the hydrodynamical model results.

\end{document}